\documentclass[journal]{IEEEtran}
\ifCLASSINFOpdf
\else
\fi
\usepackage{cite}% cite the reference
\usepackage{indentfirst}
\usepackage{graphicx}
\usepackage{amssymb}%beautiful large equal \geqslant, small equal \leqslant ; not beautiful large equal  \geq, small equal \leq
\usepackage{amsfonts}
\usepackage{amsmath}
\usepackage{cases}
\usepackage{algpseudocode}
\usepackage{subfigure}
\usepackage{mathrsfs}
\usepackage{makecell}
\usepackage[linesnumbered,boxed,ruled,commentsnumbered]{algorithm2e}
\SetKwRepeat{Do}{do}{while}%
\usepackage[ruled]{algorithm2e}
\usepackage{amsthm}
\usepackage{multirow}
\usepackage{url}
\usepackage{amssymb}
\usepackage{color}
\usepackage{textcomp}
\usepackage{afterpage}
\usepackage{amsthm} % Required for proof environments
\usepackage{environ} % Required for defining environments with \NewEnviron

\newtheoremstyle{baitingstyle}{3pt}{3pt}{\itshape}{1.2em}{\itshape}{:}{0.5em}{{\thmname{#1}\thmnumber{ #2}\thmnote{ (#3)}}}
\theoremstyle{baitingstyle}
\newtheorem{theorem}{Theorem}
\theoremstyle{plain}

\theoremstyle{baitingstyle}
\newtheorem{lemma}{Lemma}
\theoremstyle{baitingstyle}
\newtheorem{remark}{Remark}
\theoremstyle{baitingstyle}

\theoremstyle{baitingstyle}
\newtheorem{definition}{Definition}
\theoremstyle{baitingstyle}

\renewenvironment{proof}{{\indent\indent\itshape Proof of Theorem 1:}\,}{\hfill$\QEDA$}
\NewEnviron{prooflemma1}{
  \par\indent\indent\itshape Proof of Lemma 1:\quad
  \BODY
  \hfill$\QEDA$\par
}
\newcommand*{\QEDA}{\hfill\ensuremath{\blacksquare}} 

\begin{document}
%
% paper title
% Titles are generally capitalized except for words such as a, an, and, as,
% at, but, by, for, in, nor, of, on, or, the, to and up, which are usually
% Linebreaks \\ can be used within to get better formatting as desired.
% Do not put math or special symbols in the title.
\title{Large-Scale Multi-Fleet Platoon Coordination: \\A Dynamic Programming Approach}

\author{Ting~Bai,~\IEEEmembership{Member,~IEEE,}
        Alexander~Johansson, Karl~Henrik~Johansson,~\IEEEmembership{Fellow,~IEEE,}
        and~Jonas~M{\aa}rtensson,~\IEEEmembership{Member,~IEEE}% <-this % stops a space
\thanks{This work was supported in part by the Swedish Research Council Distinguished Professor under Grant 2017-01078, in part by the Knut and Alice Wallenberg Foundation, and in part by the Swedish Strategic Research Foundation CLAS under Grant RIT17-0046. The work of Ting Bai was also supported by the Outstanding Ph.D. Graduate Development Scholarship from Shanghai Jiao Tong University.}\vspace{1.5pt}% <-this % stops a space

\thanks{Ting Bai, Alexander Johansson, Karl Henrik Johansson, and Jonas M{\aa}rtensson are with the Integrated Transport Research Laboratory, the Division of Decision and Control Systems, and the Digital Futures, KTH Royal Institute of Technology, 100 44 Stockholm, Sweden (e-mail: tingbai@kth.se; alexjoha@kth.se; kallej@kth.se; jonas1@kth.se).}% <-this % stops a space
}

% note the % following the last \IEEEmembership and also \thanks - 
% these prevent an unwanted space from occurring between the last author name
% and the end of the author line. i.e., if you had this:
% 
% \author{....lastname \thanks{...} \thanks{...} }
%                     ^------------^------------^----Do not want these spaces!
%
% a space would be appended to the last name and could cause every name on that
% line to be shifted left slightly. This is one of those "LaTeX things". For
% instance, "\textbf{A} \textbf{B}" will typeset as "A B" not "AB". To get
% "AB" then you have to do: "\textbf{A}\textbf{B}"
% \thanks is no different in this regard, so shield the last } of each \thanks
% that ends a line with a % and do not let a space in before the next \thanks.
% Spaces after \IEEEmembership other than the last one are OK (and needed) as

% The paper headers
\markboth{IEEE Transactions on Intelligent Transportation Systems}%
{Shell \MakeLowercase{\textit{et al.}}: Bare Demo of IEEEtran.cls for IEEE Journals}
% *** Note that you probably will NOT want to include the author's ***
% *** name in the headers of peer review papers***
% the title area
\maketitle

\begin{abstract}
Truck platooning is a promising technology that enables trucks to travel in formations with small inter-vehicle distances for improved aerodynamics and fuel economy. The real-world transportation system includes a vast number of trucks owned by different fleet owners, for example, carriers. To fully exploit the benefits of platooning, efficient dispatching strategies that facilitate the platoon formations across fleets are required. This paper presents a distributed framework for addressing multi-fleet platoon coordination in large transportation networks, where each truck has a fixed route and aims to maximize its own fleet's platooning profit by scheduling its waiting times at hubs. The waiting time scheduling problem of individual trucks is formulated as a distributed optimal control problem with continuous decision space and a reward function that takes non-zero values only at discrete points. By suitably discretizing the decision and state spaces, we show that the problem can be solved exactly by dynamic programming, without loss of optimality. Finally, a realistic simulation study is conducted over the Swedish road network with $5,000$ trucks to evaluate the profit and efficiency of the approach. The simulation study shows that, compared to single-fleet platooning, multi-fleet platooning provided by our method achieves around $15$ times higher monetary profit and increases the CO$_2$ emission reductions from $0.4\%$ to $5.5\%$. In addition, it shows that the developed approach can be carried out in real-time and thus is suitable for platoon coordination in large transportation systems. 
\end{abstract}


\IEEEpeerreviewmaketitle

%=======Section I===================
\section{Introduction}
\IEEEPARstart{S}{afe} and efficient transportation systems are crucial for social development and economic growth. Over the past decades, the increasing global concerns about traffic jams, energy crises, and climate change have increased the urgency of developing techniques to build a green and efficient transportation system.

Truck platooning is an important technology supporting the development of more sustainable road goods transportation. It allows trucks to drive safely in formations with a small inter-vehicle gap to improve the fuel economy, due to the decreased aerodynamic drag endured by the trailing trucks. Taking advantage of the advances in wireless sensing and autonomous driving, platooning has developed rapidly since the mid-1990s, from the first study on truck automation conducted in ``Chauffeur" in the EU project T-TAP~\cite{gehring1997practical,zhang2020fuel}, and the first- and second-generation truck platooning with and without active drivers in the trailing trucks~\cite{schirrer2022energy}, to now gradually merging into commercialization in the foreseeable future~\cite{axelsson2020truck}.

As a promising intelligent transportation technology, truck platooning has drawn wide attention of governments, trucking industries, and research institutions in the past decade. Not only because of its advantages in increasing road capacity, improving traffic flow, and enhancing traffic safety, for instance, by reducing accidents incurred by overtaking and lane changes on highways, but also because of its high potential to reduce fuel consumption and CO$_2$ emissions. The field tests in~\cite{tsugawa2016review} have shown that platooning can achieve an average fuel savings of $13\%$ with $10$ m inter-vehicle gaps and savings of $18\%$ with $4.7$ m gaps. Given that road freight transportation accounts for two-thirds of all freight carried in the EU and is responsible for approximately $25\%$ of the vehicle-related carbon emissions~\cite{van2010eu,siskos2019assessing}, the total amount of fuel saving and carbon emission reduction from platooning can be substantial. In addition to the above merits, truck platooning also helps freight carriers save labor costs and alleviate the driver shortage. According to a report of the American Transportation Research Institute~\cite{costello2015truck}, driver cost is the highest operating cost for the trucking industry, accounting for up to $43\%$ of the total costs, followed by fuel. At the same time, the driver market has continued to tighten and many freight firms are struggling to hire enough drivers to meet the increasing freight demands~\cite{wang2022transportation}. Platooning technology promises a viable solution to partially address these issues. By enabling the following trucks to drive in an autonomous mode, truck platooning could reduce the working hours of drivers, enhance the operational savings of carriers, as well as alleviate the scarcity of drivers. 

To fully benefit from the advantages of platooning, trucks with diverse routes and scheduled departure times need coordination to form platoons efficiently and seamlessly in the transportation system. In contrast to platooning control at the physical level, where the target is to maintain a reliable running of platoons on the roads, high-level platoon coordination focuses on optimizing dispatching strategies to facilitate forming platoons among trucks. By coordination, trucks are able to merge into platoons en routes by adjusting, for example, their transport paths~\cite{larsson2015vehicle,meisen2008data}, velocities traveling on common route sections~\cite{larson2014distributed,van2017efficient,hoef2019predictive}, and velocities when entering the neighborhood driving area of other trucks~\cite{liang2015heavy}. Some studies consider the hub-based platoon coordination (\emph{i.e.}, platoons are formed only at hubs)~\cite{bai2022approximate,zhang2017freight,bai2023third,sokolov2017maximization}, where trucks schedule their waiting and departure times at hubs along their routes to maximize the platooning profit. To date, most of the literature on platoon coordination strategies has focused on vehicles owned by individual owners or the same fleet, referred to as single-vehicle and single-fleet platooning, respectively. In single-vehicle platooning, every truck can form platoons with each other but aims at optimizing its own profit from joining platoons. In single-fleet platooning, only trucks from the same fleet can form platoons and all trucks belonging to the same fleet have the common goal of maximizing the fleet-wide platooning profit.
\begin{figure}[!t]
    \centering
    \subfigure[Single-fleet platooning]{
    \includegraphics[width=8.4cm,height=4.3cm]{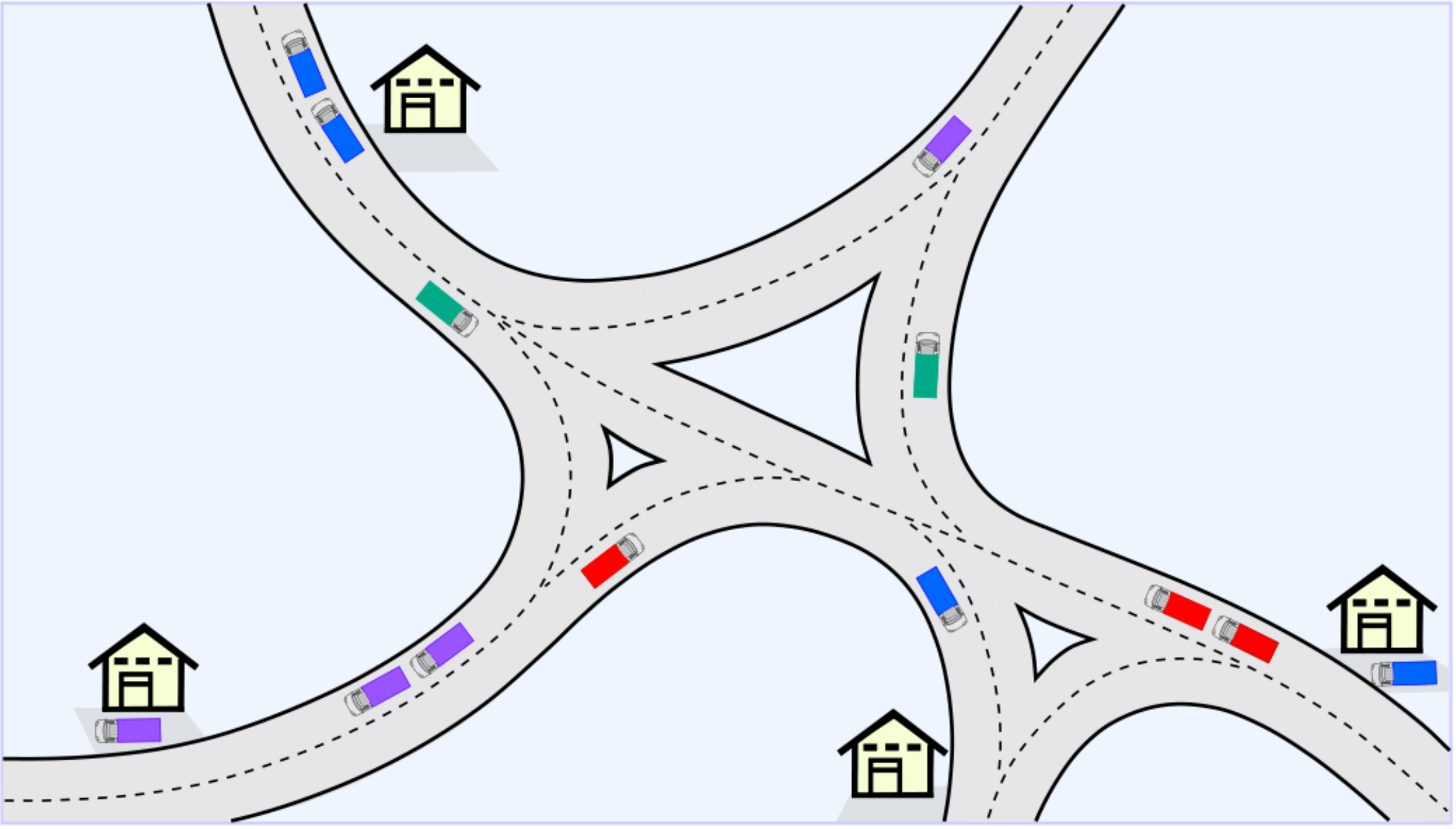}
    }
    \centering
    \subfigure[Multi-fleet platooning]{
    \includegraphics[width=8.4cm,height=4.3cm]{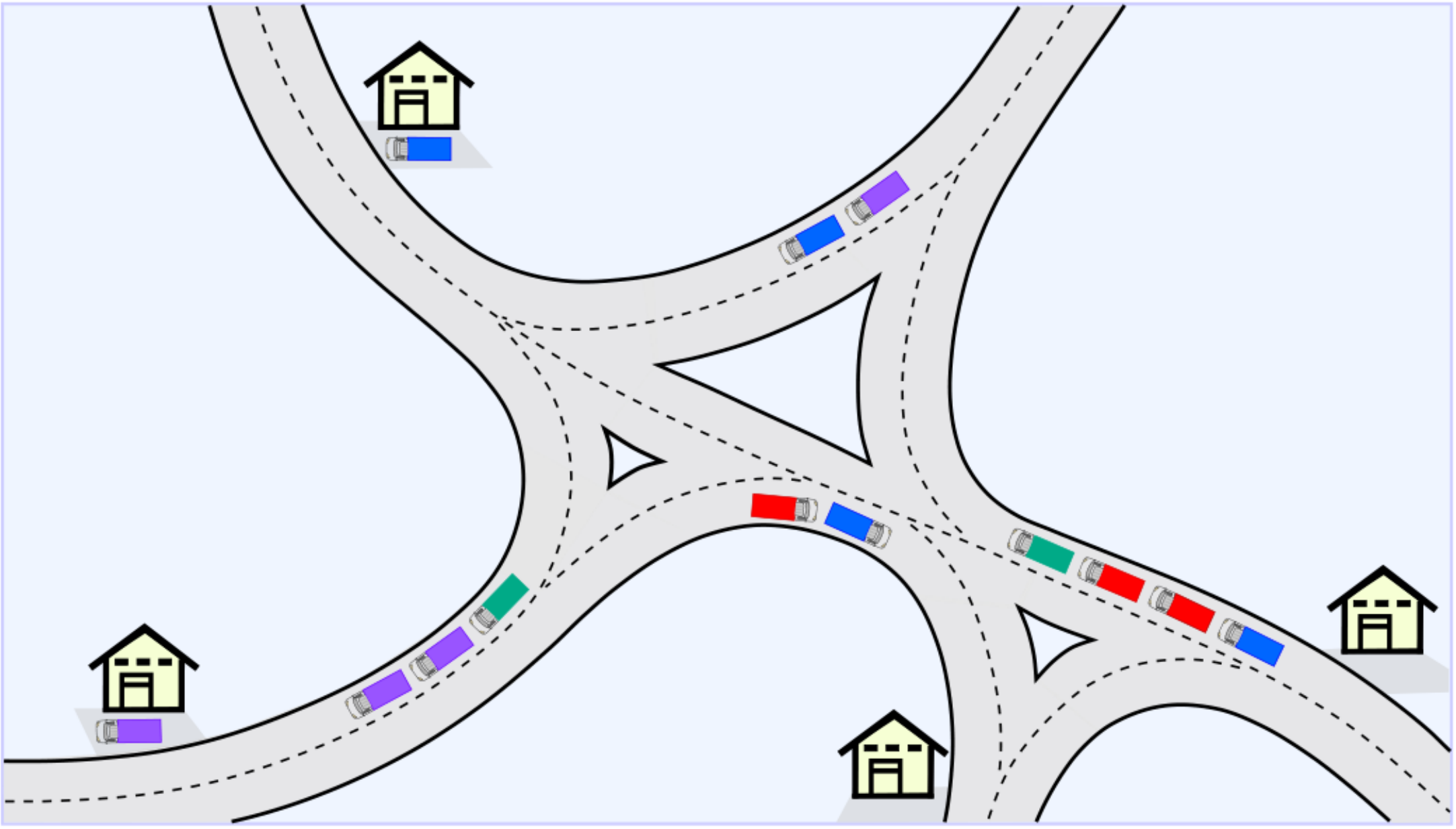}
    }
    \DeclareGraphicsExtensions.
    \caption{Single- (a) and multi-fleet (b) platooning, where trucks from the same fleet are shown in the same color. Hubs along the routes are shown by yellow houses, at which trucks can wait and form platoons with others.}
    \label{Fig.1}
\end{figure}

In many real-world applications, the fleet composition of a transportation system could be diverse, making the platoon coordination not fit in existing platooning frameworks. For example, among the $336,000$ heavy-duty trucks registered in Texas, $38\%$ of the trucks were owned by small firms with no more than $30$ trucks in each firm and $24\%$ of them were from firms with $10$ or fewer trucks~\cite{thornton2008compliance}. A similar situation can also be seen in European countries, where the trucking industry itself is highly fragmented and most of the fleets are relatively small~\cite{kubavnova2021analysis}. For these scenarios, traditional single-fleet platooning methods become inapplicable, as they are not able to coordinate trucks from different fleets even if trucks are close by. As illustrated in Figure~\ref{Fig.1}(a), single-fleet platooning coordinating trucks within the same fleet causes a loss of platooning opportunities, especially in large road networks with a vast number of trucks from different fleets. Different from single-fleet platooning, multi-fleet platooning allows trucks to cooperate across fleets for forming platoons while optimizing the profit of individual fleets. See Figure~\ref{Fig.1}(b) for an illustration of the consequence of multi-fleet platooning. 

Over the past few years, efficient platoon coordination strategies have been actively researched with some related projects involving SARTRE~\cite{robinson2010operating}, COMPANION~\cite{eilers2015companion}, and ENSEMBLE~\cite{ENSEMBLE}. Notwithstanding, studies on multi-fleet platoon coordination are still scarce, especially for dealing with large-scale platooning systems. In~\cite{caiazzo2021distributed}, the authors study the multiplatoon management problem by proposing a cooperative coordination approach for merging/splitting multiple platoons that approach road section restrictions. Therein, the objective is to ensure safe and robust formations of multiple platoons, while not considering the scheduling optimization of multi-fleet platoons. In a similar context, the authors in~\cite{vcivcic2021coordinating} address bottleneck decongestion by developing optimal control strategies to manipulate the speed and formation of platoons. Their research aims to improve throughput on road segments and reduce the total time spent by vehicles, without involving platoon scheduling optimization across fleets.  

Recently, the research efforts in \cite{johansson2021strategic} and \cite{zeng2020distributed} model the strategic interaction among trucks and carriers as non-cooperative games and consider Nash equilibria as multi-fleet platoon coordination solutions. This kind of method suffers from the inefficiency in seeking an equilibrium solution to the problem by iterations among trucks. In~\cite{johansson2021real}, the authors propose a Pareto-improving multi-fleet platoon coordination solution, where the fleet owners leave the coordination to local coordinators at hubs and each fleet owner is better off in the multi-fleet solution than by performing single-fleet platoon coordination. In~\cite{7437386}, a three-layered approach for controlling individual vehicles is proposed, where the fleet management layer focuses on transport planning and routing while the cooperation layer deals with platoon coordination. However, only spontaneous platoon formations are considered. Different from these optimization methods taking one local hub as the platoon coordinator, the method proposed in this paper takes into account the fleet's platooning profit at all hubs along trucks' routes, which is more beneficial for trucks to make a long-term plan when scheduling their waiting times and to catch up the platooning opportunities at every hub. In our previous research~\cite{bai2021event}, an event-triggered model predictive control method for platoon coordination is proposed, which enables trucks to integrate platooning opportunities at multiple hubs, whereas only single-vehicle platooning is studied.  

In this paper, we develop a distributed multi-fleet platoon coordination strategy suitable for handling large-scale transportation networks. Specifically, we consider hub-based platoon coordination, where trucks are owned by different fleet owners and have fixed transport routes in the road network. Every truck allows for forming platoons with others but has the target of maximizing its own fleet's platooning profit by optimally scheduling its waiting times at hubs. The waiting time scheduling problem of individual trucks is formulated as a distributed optimal control problem that has a continuous decision space and a reward function taking non-zero values only at discrete points. We show that the decision and state spaces of the problem can be discretized, and propose an optimal solution to the problem based on dynamic programming (DP). In our recent work~\cite{9976309}, an intuitive illustration of the platoon coordination scheme proposed in this paper is presented, without getting into intricate technical details. The contributions of this paper are summarized as follows. 
\begin{itemize}
    \item We formulate the hub-based multi-fleet platoon coordination problem as a distributed optimal control problem, where the reward function captures the platooning reward that each truck provides to its fleet by forming platoons and the loss caused by waiting at hubs.      
    \vspace{1.5pt}
    
    \item We show that the optimal solution of the formulated problem lies in a discrete decision space, which allows us to discretize the decision and state spaces and then solve the problem by DP, without loss of optimality.  
    \vspace{1.5pt}
    
    \item We propose a DP-based optimal solution to the problem and present algorithms to compute the optimal waiting time decisions, building upon the discretized decision and state spaces. The reduced search space contributes to an efficient solution for addressing the problem.  
    \vspace{1.5pt}

    \item  We perform a realistic simulation study over the Swedish road network with $5,000$ trucks, where the trips and fleet size distribution are generated from real data. The simulation study shows that the proposed multi-fleet platoon coordination method achieves about $15$ times higher profit and $13$ times higher CO$_2$ emission reduction compared to the single-fleet platooning. The simulation study also shows that the developed DP-based solution approach can be carried out in real-time when handling large transportation networks.  
\end{itemize}

The rest of the paper is organized as follows. Section~\ref{Section II} introduces the road network, truck dynamics, and the problem formulation of the multi-fleet platoon coordination. In Section~\ref{Section III}, the modeling of the stage reward function and the terminal reward function in our problem is introduced, respectively. Section~\ref{Section IV} presents a DP-based optimal solution for solving the multi-fleet platoon coordination problem. Section~\ref{Section V} gives the simulation results that demonstrate the profit and efficiency of the developed approach. Eventually, Section~\ref{Section VI} concludes this paper and outlines the directions to investigate in future work. 

%==============Section II=====================
%------------Problem Formulation--------------
\section{Problem Formulation}\label{Section II}
%%%%%%%%%%%%%%%%%%%%%%%%%%%%%%%%%%%%%
%Section II, Subsection A {Road Network}
This section introduces the problem formulation of the hub-based multi-fleet platoon coordination, including the considered road network, the dynamic models of individual trucks, and the multi-fleet platoon coordination problem. 
 
\subsection{Road Network}
Consider a road network defined over a directed graph~$\mathcal{G}=\!(\mathcal{H},\mathcal{E})$ with a node set $\mathcal{H}$ and an edge set $\mathcal{E}$. Each node in $\mathcal{H}$ represents a hub in the road network and each edge in~$\mathcal{E}$ represents a road segment connecting a pair of hubs. We consider $M$ trucks with fixed transport routes in the road network $\mathcal{G}$, and each of the trucks belongs to one of the fleets in the set $\mathcal{F}\!=\!\{\mathcal{F}_1,\dots,\mathcal{F}_S\}$.  
To simplify the presentation of the problem, we assume that each truck starts its trip at one hub and ends its trip at another hub. For any truck $i\!\in\!\mathcal{M}\!=\!\{1,\dots,M\}$, the starting hub and the destination hub of its trip are referred to as its first and $N_i$-th hub, respectively. The route of truck $i$ is then denoted by
\begin{align}
    \mathcal{E}_i=\big\{\textbf{\textit{e}}_{i,1},\dots,\textbf{\textit{e}}_{i, N_i-1}\big\},\nonumber
\end{align}
where $\textbf{\textit{e}}_{i,k}$ represents the $k$-th road segment in the route of truck $i$. More specifically, it is the road segment connecting the $k$-th and $(k\!+\!1)$-th hub in truck $i$'s route.
  
%%%%%%%%%%%%%%%%%%%%%%%%%%%%%%%%%%%%%
%Section II, Subsection B {Truck Dynamics}
\subsection{Truck Dynamics}
In a hub-based platooning system, trucks make their waiting time decisions each time arriving at a hub to form platoons with other trucks. The dynamics of any truck $i$ can be characterized by
\begin{align}
    t^a_{i,k+1}&=f_{i,k}(t_{i,k}^a,t_{i,k}^w),~~k\!=\!1,\dots{N_i}\!-\!1,\nonumber
\end{align}
where $t_{i,k}^a$ is the state of truck $i$, denoting its arrival time at its $k$-th hub, and $t_{i,k}^w$ is the waiting time decision of truck $i$ at its $k$-th hub. Specifically, $f_{i,k}(t_{i,k}^a,t_{i,k}^w)$ has the following form
\begin{align}
    f_{i,k}(t_{i,k}^a,t_{i,k}^w)=t^a_{i,k}\!+t^w_{i,k}\!+\tau(\textbf{\textit{e}}_{i,k}),\label{Eq.1}
\end{align}
where $\tau(\textbf{\textit{e}}_{i,k})$ denotes the travel time required for truck $i$ to traverse its $k$-th road segment. In line with the state transition function $f_{i,k}(t_{i,k}^a,t_{i,k}^w)$, truck $i$ can update its state from its $k$-th hub to its $(k\!+\!1)$-th hub. We denote by $t_{i,1}^a\!=\!t_i^0$ the arrival time of truck $i$ at its starting hub. Moreover, we assume that every truck has a delivery deadline to respect at its destination hub, which requires that $t_{i,N_i}^a\!\!\leq\!{t_i^{dd}}$, where $t_i^{dd}$ represents the delivery deadline of truck $i$ at its destination hub. 

\begin{remark}{\label{Remark1}}
This paper considers a simplified truck dynamic model in (\ref{Eq.1}), where trucks' travel times on roads $\tau(\textbf{\textit{e}}_{i,k})$ are assumed to be deterministic. In real transportation networks, however, trucks' travel times can be uncertain if, for example, considering traffic congestion. As travel time uncertainty is not the focus of this paper, we have not included it in our method. Nevertheless, this uncertainty could be incorporated by allowing trucks to update their predictions of arrival times at hubs periodically.
\end{remark}
%%%%%%%%%%%%%%%%%%%%%%%%%%%%%%%%%%%%%
%Section II, Subsection C {Platoon Coordination Problem}
\subsection{Platoon Coordination Problem}
\begin{figure}[t]
     \centering
     \includegraphics[width=1.0\linewidth]{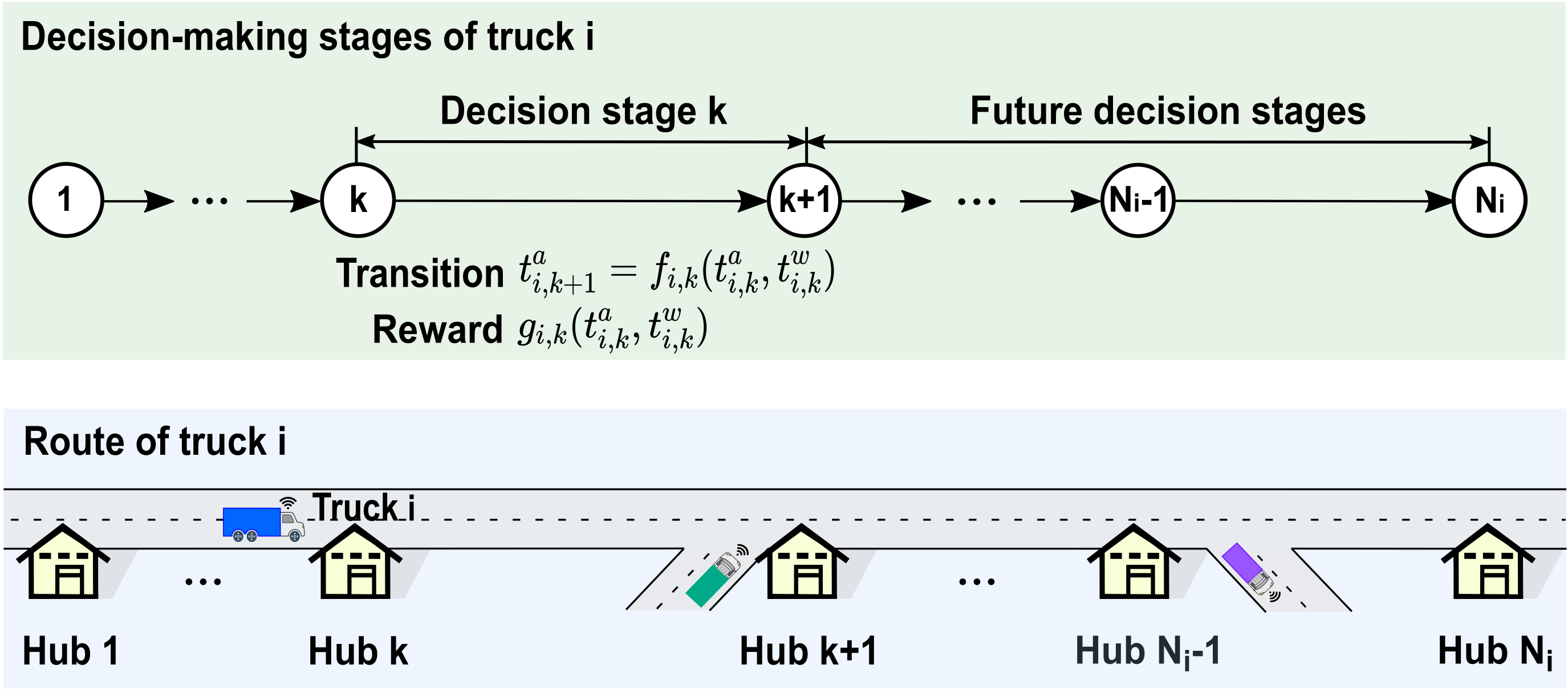}
      \caption{Decision-making stages of truck $i$, which travels from its first hub to its $N_i$-th hub to fulfill a delivery task. Truck $i$ makes its waiting time decisions each time arriving at a hub in its route.}
      \label{Fig.2}
\end{figure}

For any truck $i$ arriving at its $k$-th hub, the aim is to maximize the increased platooning reward that truck $i$ provides to its fleet. This is achieved by optimizing its waiting time decisions at every hub in its remaining route, \emph{i.e.}, from its $k$-th hub to its destination hub. As illustrated in Figure~\ref{Fig.2}, truck $i$ makes its waiting time decisions $\{t_{i,k}^w,\dots,t_{i,N_i-1}^w\}$ when arriving at its $k$-th hub. Given the arrival time $t_{i,k}^a$, the optimization problem solved by truck $i$ at its $k$-th hub can be formulated as 
\begin{align}
    &\!\!\!J_{i,k}^{*}\big(t_{i,k}^a\big)\nonumber\\
    &\!\!\!=\max\limits_{\substack{t_{i,m}^w\in\Gamma_{i,m}(t_{i,m}^a),\\t_{i,m+1}^a=f_{i,m}(t_{i,m}^a,t_{i,m}^w),\\m=k,\dots,N_i-1}}\!\!\!\!\!\!g_{i,N_i}(t_{i,N_i}^a)+\!\sum_{m=k}^{N_i-1}g_{i,m}(t_{i,m}^a,t_{i,m}^w),\label{Eq.2}
\end{align}
where $g_{i,m}(t_{i,m}^a,t_{i,m}^w)$ denotes the increased platooning reward that truck $i$ provides to its fleet by determining its waiting time at its decision stage $m$ (called as \textit{stage reward function}), and $g_{i,N_i}(t_{i,N_i}^a)$ denotes the terminal reward incurred at the destination hub, referred to as \textit{terminal reward function}. To follow the delivery deadline, the waiting time at each hub $m\!=\!k,\dots,N_i\!-\!1$ is restricted by $t_{i,m}^w\!\in\!\Gamma_{i,m}(t_{i,m}^a)$ with
\begin{align}
    \!\!\!\Gamma_{i,m}(t_{i,m}^a)\!=\!\Bigg\{t_{i,m}^w\,\bigg|\, {0}\leq{t_{i,m}^w}\!\leq{t_i^{dd}\!-t_{i,m}^a}\!-\!\!\!\sum_{m=k}^{N_i-1}\!\!\tau{(\textbf{\textit{e}}_{i,m})}\!\Bigg\},\label{Eq.3}
\end{align}
where, as previously defined, $t_i^{dd}$ is the delivery deadline of truck $i$. The upper bound of $t_{i,m}^w$ in (\ref{Eq.3}) is the longest waiting time that truck $i$ can use at its $m$-th hub without waiting at all the following hubs, which depends on truck $i$'s delivery deadline, its arrival time at its $m$-th hub, and the time needed to travel from its $m$-th hub to the destination hub. 

By addressing the optimization problem formulated in (\ref{Eq.2}), truck $i$ can determine its optimal waiting time sequence $\{t_{i,k}^{w*},\dots,t_{i,N_i-1}^{w*}\}$ at all the remaining hubs, where $t_{i,k}^{w*}$ computed for its $k$-th hub will be implemented in its real control, while the rest waiting times $\{t_{i,k+1}^{w*},\dots,t_{i,N_i-1}^{w*}\}$, and the corresponding arrival times at the following hubs will be used by other trucks as truck $i$'s predicted departure times at hubs when optimizing their waiting times. 

%==========Section III=====================================
%------Multi-fleet Platooning Reward Modeling-------------
\section{Multi-fleet Platooning Reward Modeling}\label{Section III}
This section proposes the modeling of the stage reward function $g_{i,m}(t_{i,m}^a,t_{i,m}^w)$ that captures the fleet's increased platooning reward in multi-fleet platoon coordination, and the modeling of the terminal reward function $g_{i,N_i}(t_{i,N_i}^a)$ that captures the loss each truck causes to its fleet due to waiting. We start by introducing the conditions trucks need to meet to form a platoon at a hub. On this basis, the models of the stage and terminal reward functions are presented, respectively.

%%%%%%%%%%%%%%%%%%%%%%%%%%%%%%%%%%%%%
%Section III, Subsection A {Conditions to Form Platoons}
\subsection{Conditions to Form Platoons}
Towards forming a platoon with other trucks at the $k$-th hub of truck $i$, the following two conditions need to be met. 

\noindent (i) Truck $i$ has its $k$-th road segment in common with others. 

\noindent (ii) Truck $i$ leaves its $k$-th hub at the same time as other trucks. 

We refer to the other trucks satisfying the first condition as the potential platoon partners of truck $i$ at its $k$-th hub and name the trucks that meet both the two conditions as its predicted platoon partners at the $k$-th hub. Mathematically, the potential and predicted platoon partners are defined below.        
\begin{definition}\label{Definition 1}
(Potential Platoon Partner): For any truck $j\!\in\!\mathcal{M}\!\setminus\!{\{i\}}$, if truck $j$ has the $k$-th road segment of truck $i$ in its route, we say truck $j$ is a potential platoon partner of truck $i$ at its $k$-th hub. The set of potential platoon partners of truck $i$ at its $k$-th hub is denoted by
\begin{align}
 \mathcal{P}_{i,k}=\Big\{j\,\big|\, \textbf{\textit{e}}_{i,k}\!\in\!\mathcal{E}_j,~ j\!\in\!\mathcal{M}\!\setminus\!{\{i\}}\Big\}.\nonumber
\end{align}
\end{definition}

\begin{definition}\label{Definition 2}
(Predicted Platoon Partner): 
Given truck $i$'s arrival time $t_{i,k}^a$ and the predicted departure times $t_{i,k}^{d,j}$ of other trucks $j\!\in\!\mathcal{P}_{i,k}$, we refer to trucks in $\mathcal{P}_{i,k}$ that truck $i$ predicts to form a platoon with at its $k$-th hub if applying the waiting time $t_{i,k}^w$ as its predicted platoon partners at its hub $k$, which are trucks in the set 
\begin{align}
\mathcal{R}_{i,k}(t_{i,k}^a,t_{i,k}^w)=\Big\{j\!\in\!\mathcal{P}_{i,k}\,\big|\, t_{i,k}^a\!+t_{i,k}^w\!=t_{i,k}^{d,j}\Big\}.\label{Eq.4}
\end{align}
\end{definition}

Notice that the predicted platoon partner set $\mathcal{R}_{i,k}(t_{i,k}^a,t_{i,k}^w)$ is a function of the arrival time $t_{i,k}^a$ and the waiting time $t_{i,k}^w$, meaning that truck $i$ can decide its predicted platoon partners at its $k$-th hub by controlling its waiting time $t_{i,k}^w$. 

By Definition~\ref{Definition 1}, with the knowledge of the routes of other trucks, every truck can compute offline the set of its potential platoon partners $\{\mathcal{P}_{i,k},\dots,\mathcal{P}_{i, N_i-1}\}$ at each of its hubs. A truck in $\mathcal{P}_{i,k}$ means that truck $i$ will consider this truck as a potential platoon partner at its $k$-th hub to form platoons with. Additionally, given the predicted departure times of other trucks in the set $\mathcal{P}_{i,k}$, truck $i$ is able to determine optimally its predicted platoon partner set $\mathcal{R}_{i,k}(t_{i,k}^a,t_{i,k}^w)$ online by optimizing its waiting time $t_{i,k}^w$ such that the resulting platooning reward to its fleet is maximized. 

%%%%%%%%%%%%%%%%%%%%%%%%%%%%%%%%%%%%%
%Section III, Subsection B {Stage Reward Function}
\subsection{Stage Reward Function Modeling}
Based on the above conditions for forming platoons at hubs, the modeling of the stage reward function $g_{i,m}(t_{i,m}^a,t_{i,m}^w)$ is proposed in this subsection, which is used to characterize the increased platooning reward that truck $i$ provides to its fleet due to its waiting time decision at its $m$-th hub, where $m\!=\!k,\dots,N_i\!-\!1$.

Forming a platoon at the $m$-th hub of truck~$i$ enables truck~$i$ to enjoy the platooning benefit gained on its road segment $\textbf{\textit{e}}_{i,m}$ and thus increases its fleet's platooning profit. To depict the increased fleet platooning profit resulting from truck $i$'s waiting time decision at its $m$-th hub, first, the predicted platoon partners of truck $i$ at its $m$-th hub are distinguished. Recall that the predicted platoon partner set $\mathcal{R}_{i,m}(t_{i,m}^a,t_{i,m}^w)$ represents the set of other trucks that truck $i$ predicts to form a platoon with at its $m$-th hub. We assume that truck $i$ belongs to the fleet $\mathcal{F}_s\!\in\!\mathcal{F}$, and use $p_{i,m}^s$ and $p_{i,m}^{-s}$ to denote the number of the predicated platoon partners in $\mathcal{R}_{i,m}(t_{i,m}^a,t_{i,m}^w)$ that belong to the same fleet as truck $i$ and belong to a different fleet from truck $i$, respectively. More precisely, we define
\begin{subequations}
\begin{align}
    &p_{i,m}^s(t_{i,m}^a,t_{i,m}^w)=\big|\mathcal{R}_{i,m}(t_{i,m}^a,t_{i,m}^w)\!\cap\!{\mathcal{F}_s}\big|\label{Eq.5a}\\
   &p_{i,m}^{-s}(t_{i,m}^a,t_{i,m}^w)=\big|\mathcal{R}_{i,m}(t_{i,m}^a,t_{i,m}^w)\!\setminus\!{\mathcal{F}_s}\big|.\label{Eq.5b}
\end{align}
\end{subequations}

To proceed, the increased platooning profit that truck $i$ provides to its fleet $\mathcal{F}_s$ due to its waiting time decision at its $m$-th hub is introduced. According to the field experiments on truck platooning, see, \emph{e.g.}, \cite{davila2013environmental,bishop2017evaluation}, each following truck in a platoon has approximately the same fuel savings while the lead truck has significant small fuel savings. In this paper, we assume that every following truck has the same platooning benefit, the lead truck has zero benefits from platooning, and all trucks in the platoon evenly share the achieved platooning profit. Therefore, for a platoon of size $n$ (\emph{i.e.}, the number of trucks included in the platoon) traveling on the road segment $\textbf{\textit{e}}_{i,m}$, the average platooning profit achieved by each truck in the platoon is denoted by
\begin{align}
    \xi{\tau}(\textbf{\textit{e}}_{i,m})\frac{n\!-\!1}{n},\nonumber
\end{align}
where $\xi$ represents the monetary platooning profit per travel time unit per following truck, and the number of following trucks in the platoon is $(n\!-\!1)$. Taking this as a basis, if truck $i$ decides not to form a platoon with its predicted platoon partners in $\mathcal{R}_{i,m}(t_{i,m}^a,t_{i,m}^w)$, the platooning profit that its fleet gains from the trucks in $\mathcal{R}_{i,m}(t_{i,m}^a,t_{i,m}^w)$ that belong to the same fleet as truck $i$ is 
\begin{align}
    F_{i,m}^p=\xi{\tau}(\textbf{\textit{e}}_{i,m})\frac{p_{i,m}^s\!+\!p_{i,m}^{-s}\!-\!1}{p_{i,m}^s\!+\!p_{i,m}^{-s}}p_{i,m}^s,\nonumber
\end{align}
where $n\!=\!p_{i,m}^s\!+\!p_{i,m}^{-s}$ is the size of the platoon formed by the trucks in the predicted platoon partner set of truck $i$ at its $m$-th hub, and as defined in (\ref{Eq.5a}), $p_{i,m}^s$ trucks are from the same fleet as truck $i$ in the formed platoon. 

If truck $i$ decides to form a platoon with its predicted platoon partners at its $m$-th hub, the platooning profit that its fleet gains on the road segment $\textbf{\textit{e}}_{i,m}$ is
\begin{align}
    \Hat{F}_{i,m}^p=\xi{\tau}(\textbf{\textit{e}}_{i,m})\frac{p_{i,m}^s\!+\!p_{i,m}^{-s}}{p_{i,m}^s\!+\!p_{i,m}^{-s}\!+\!1}(p_{i,m}^s\!+\!1),\nonumber
\end{align}
where the size of the platoon is increased by $1$ as truck $i$ joins the platoon. Thus, the increased platooning profit that truck $i$ provides to its fleet by forming a platoon with its predicted platoon partners at its $m$-th hub is
\begin{align}
    \Delta F_{i,m}^p(p_{i,m}^s,p_{i,m}^{-s})&=\Hat{F}_{i,m}^p-F_{i,m}^p\nonumber\\
    & = \xi{\tau}(\textbf{\textit{e}}_{i,m})\Delta f(p_{i,m}^s,p_{i,m}^{-s}),\label{Eq.6}
\end{align}
where the increased fleet platooning profit $\Delta F_{i,m}^p(p_{i,m}^s,p_{i,m}^{-s})$ is a function of $p_{i,m}^s$ and $p_{i,m}^{-s}$. In line with the definitions in (\ref{Eq.5a}) and (\ref{Eq.5b}), $\Delta f(p_{i,m}^s,p_{i,m}^{-s})$ is of the form 
\begin{align}
&\!\!\!\!\Delta{f}(p_{i,m}^s,p_{i,m}^{-s})=\nonumber\\
&\!\!\!
\begin{cases}
    \!\!1\!-\!\frac{p_{i,m}^{-s}}{\big(p_{i,m}^s+p_{i,m}^{-s}+1\big)\big(p_{i,m}^s+p_{i,m}^{-s}\big)}, & \!\!\!\text{if $\mathcal{R}_{i,m}(t_{i,m}^a,t_{i,m}^w)\!\neq\!{\emptyset}$}\\
    \!\!0, & \!\!\!\text{otherwise}.\label{Eq.7}
    \end{cases}
\end{align}

We note that given the arrival time $t_{i,m}^a$ and the predicted departure times of other trucks in $\mathcal{P}_{i,m}$, truck $i$ is able to regulate its waiting time $t_{i,m}^w$ to form a platoon with different predicted platoon partners. By~(\ref{Eq.4}), (\ref{Eq.6}) and (\ref{Eq.7}), it can be seen that different waiting times correspond to different predicted platoon partner sets and result in different increased fleet platooning profits. This indicates that truck $i$ can select optimally its platoon partners in its predicted platoon partner set by optimizing its waiting time such that its fleet's platooning profit is maximally increased.  

Given the above, the stage reward function of truck~$i$ on its decision stage $m\!=\!k,\dots,N_i\!-\!1$ can be modeled by
\begin{align}
    &\!\!\!\!g_{i,m}(t_{i,m}^a,t_{i,m}^w)\nonumber\\
    &\quad \quad =\Delta F_{i,m}^p(p_{i,m}^{s},p_{i,m}^{-s})\nonumber\\
    &\quad \quad =\xi{\tau}(\textbf{\textit{e}}_{i,m})\Delta f\big(p_{i,m}^s(t_{i,m}^a,t_{i,m}^w),p_{i,m}^{-s}(t_{i,m}^a,t_{i,m}^w)\big),\label{Eq.8}
\end{align}
which captures the increased platooning reward of the fleet $\mathcal{F}_s$ contributed by truck $i$'s waiting time decision at its $m$-th hub.

%%%%%%%%%%%%%%%%%%%%%%%%%%%%%%%%%%%%%%%%%%%%
\subsection{Terminal Reward Function Modeling}
Trucks' waiting time decisions at hubs can increase their fleet's platooning profit, but may also increase the waiting loss of the fleet due to the additional labor costs and the higher risk of being delayed. In this paper, the waiting loss that truck $i$ causes to its fleet is captured by its terminal reward function using the following form
\begin{align}
    g_{i,N_i}(t_{i,N_i}^a)=-\epsilon_i\Bigg(t_{i,N_i}^a\!-\!t_{i,k}^a-\!\!\sum_{m=k}^{N_i-1}\tau{(\textbf{\textit{e}}_{i,m})}\!\Bigg),\label{Eq.9}
\end{align}
where $\epsilon_i$ represents the monetary loss that truck $i$ causes to its fleet per time for waiting. Truck $i$'s total waiting time at all hubs in its remaining route is denoted by $t_{i,N_i}^a\!-t_{i,k}^a\!-\sum_{m=k}^{N_i-1}\tau(\textbf{\textit{e}}_{i,m})$. The waiting loss modeled in (\ref{Eq.9}) shows that the later a truck reaches its destination hub, the higher the waiting loss its fleet will sustain. Moreover, it is worth noting that the value of $\epsilon_i$ can be varied for trucks with different labor costs or penalties for being delayed. For instance, a truck that transports food bears a higher loss per waiting time than a truck transporting clothing, which leads to a larger $\epsilon_i$ in the terminal reward function.

At this point, the hub-based multi-fleet platoon coordination problem has been presented completely in (\ref{Eq.2}), where the stage reward function $g_{i,m}(t_{i,m}^a,t_{i,m}^w)$ is modeled in (\ref{Eq.8}) while the terminal reward function $g_{i,N_i}(t_{i,N_i}^a)$ is modeled in (\ref{Eq.9}). By solving the problem (\ref{Eq.2}) under the constraint (\ref{Eq.3}), the optimal waiting time decisions of each truck can be obtained.  

\begin{remark}\label{Remark2}
The multi-fleet platoon coordination problem formulated in (\ref{Eq.2}) is hard to solve due to the structure of the stage reward function in (\ref{Eq.8}), which is non-differentiable and thus cannot be addressed by gradient methods. Additionally, the decision space $\Gamma_{i,m}(t_{i,m}^a)$ of the problem given in (\ref{Eq.3}) is continuous and infinite, making it difficult to apply DP with uniform discretization to solve the problem.      
\end{remark}

%==========Section IV====================================
%------Dynamic Programming Solution-------------
\section{Dynamic Programming Solution}\label{Section IV}
This section presents a DP-based optimal solution for solving the problem formulated in (\ref{Eq.2}). To this end, we first formulate the Bellman optimality equation (BOE) used to compute the optimal solution and show that the decision space of the problem can be discretized, without loss of optimality. Then, we will show how to generate the discrete state space using the discrete decision space. Finally, we explain how to find the optimal solution to the problem in (\ref{Eq.2}) by applying the DP method using the discrete decision and state spaces.

%%%%%%%%%%%%%%%%%%%%%%%%%%%%%%%%%%%%%
%Section IV, Subsection A {Bellman Optimality Equation}
As is known, a long-term reward function in a given control sequence can be decomposed into the reward function in the current control action, and the reward function in the future actions, formulated by the BOE. To compute the optimal waiting times in problem (\ref{Eq.2}), we define that
\begin{align}
    J_{i,N_i}^{*}\big(t_{i,N_i}^a\big)=g_{i,N_i}(t_{i,N_i}^a),\quad \quad \text{for all}~t_{i,N_i}^a,\label{Eq.10}
\end{align}
and for each decision stage $m\!=\!k,\dots,N_i\!-\!1$, we formulate the BOE as  
\begin{align}
    J_{i,m}^{*}\big(t_{i,m}^a\big)\!=\!\max\limits_{t_{i,m}^w\in{\Gamma_{i,m}(t_{i,m}^a)}}Q_{i,m}\big(t_{i,m}^a,t_{i,m}^w\big),\label{Eq.11}
\end{align}
where
\begin{align}
&\!\!\!\!Q_{i,m}\big(t_{i,m}^a,t_{i,m}^w\big)\nonumber\\
&\quad \quad \quad =\!g_{i,m}(t_{i,m}^a,t_{i,m}^w)+J_{i,m+1}^{*}\big(f_{i,m}(t_{i,m}^a,t_{i,m}^w)\big),\label{Eq.12}
\end{align}
where the first term in (\ref{Eq.12}) is the reward at the decision stage $m$ and the second term is the rewards at the future decision stages. By solving the BOE in~(\ref{Eq.11}), one can obtain the optimal waiting time decisions for truck $i$ at each of its decision stages. However, as the decision and state spaces associated with the problem (\ref{Eq.2}) are continuous, while the stage reward function modeled in (\ref{Eq.8}) takes non-zero values only at discrete points, this BOE is hard to solve. 

In order to solve the BOE in~(\ref{Eq.11}) by DP, the following lemma is proposed, which shows that the continuous decision space of the problem can be discretized while not losing the solution optimality. Later, we will use the discretized decision space to generate the discrete state space and show how to solve the problem (\ref{Eq.2}) by DP.   
\IncMargin{0.57em}  
\begin{algorithm}[t]
\SetKwInOut{Input}{\textbf{Input}}\SetKwInOut{Output}{\textbf{Output}} % 替换关键词
    \Input{The arrival time $t_{i,k}^a$, the predicted departure times $t_{i,m}^{d,j}$ of trucks $j\!\in\!\mathcal{P}_{i,m}$ for $m\!=\!k,\dots,N_i\!-\!1$.}
    
    \Output{The discrete state space $\Gamma_{i,m}^S$ associated with the problem (\ref{Eq.2}) for $m\!=\!k,\dots,N_i$.}
    \BlankLine
    {Set $\Gamma_{i,k}^S\!=\!\{t_{i,k}^a\}$\;}\vspace{3pt}
    
    \For{$m\!=\!k,\dots,N_i\!-\!1$}{\vspace{3pt} \For{$t_{i,m}^a\!\in\!\Gamma_{i,m}^S$}{  
  \vspace{3pt}
  Compute $\Gamma_{i,m}^D(t_{i,m}^a)$ by (\ref{Eq.13})\;
  } 
  \vspace{2pt}
  Compute the discrete state space\\
  \vspace{2pt}
  $\Gamma_{i,m+1}^S\!=$\\
  $\Big\{f_{i,m}(t_{i,m}^a,t_{i,m}^w)\,\big|\, t_{i,m}^a\!\in\!{\Gamma_{i,m}^S},t_{i,m}^w\!\in\!{\Gamma_{i,m}^D}(t_{i,m}^a)\Big\}$\;
  \vspace{1pt}
  }\vspace{1pt}
  \textbf{return} $\Gamma_{i,m}^S$ for $m\!=\!k,\dots,N_i$.
  \caption{Generate Discrete State Space}\label{Alg.1}
\end{algorithm}

\begin{lemma}\label{Lemma1}
The optimal value function $J_{i,m}^{*}\big(t_{i,m}^a\big)$ in (\ref{Eq.11}) can be achieved by solving the following BOE 
\begin{align}
J_{i,m}^{*}\big(t_{i,m}^a\big)=\max\limits_{t_{i,m}^w\in{\Gamma_{i,m}^D(t_{i,m}^a)}}Q_{i,m}\big(t_{i,m}^a,t_{i,m}^w\big),\nonumber
\end{align}
where $\Gamma_{i,m}^D(t_{i,m}^a)$ denotes the discrete decision space with respect to the arrival time $t_{i,m}^a$, and is defined as
\begin{align}
    &\!\!\!\Gamma_{i,m}^D(t_{i,m}^a)\nonumber\\
    &\quad \quad \quad~ =\Big\{t_{i,m}^{d,j}-t_{i,m}^a\!\in\!\Gamma_{i,m}(t_{i,m}^a)\,\big|\,j\!\in\!\mathcal{P}_{i,m}\Big\}\!\cup\!\big\{0\big\},\label{Eq.13}
\end{align}
where, as previously defined, $t_{i,m}^{d,j}$ is the predicted departure time of a potential platoon partner $j$ of truck $i$ at its hub $m$.
\vspace{1pt}
\end{lemma}
\begin{prooflemma1}
See the Appendix.
\end{prooflemma1}
\vspace{2pt}

Lemma~\ref{Lemma1} shows that the optimal solution to the BOE in~(\ref{Eq.11}) lies in the discrete decision space $\Gamma_{i,m}^D(t_{i,m}^a)$. In line with this discrete decision space and the state transition function $f_{i,m}(t_{i,m}^a,t_{i,m}^w)$ in (\ref{Eq.1}), the corresponding discrete state space can be initialized and generated recursively, as given in Algorithm~\ref{Alg.1}. 

\begin{remark}\label{Remark3}
In the problem (\ref{Eq.2}), although trucks' arrival and departure times at hubs are continuous, we demonstrate in the proof of Lemma~\ref{Lemma1} that it is never optimal for trucks to wait at a hub unless a platoon is joined. In other words, waiting time decisions that are not in the discrete decision space defined in (\ref{Eq.13}) are redundant when seeking the optimal solution, which makes our solution optimal for the problem.
\end{remark}
%%%%%%%%%%%%%%%%%%%%%%%%%%%%%%%%%%%%%
%Section IV, Subsection B {Dynamic Programming Solution}
\IncMargin{0.57em}  
\begin{algorithm}[t]
\SetKwInOut{Input}{\textbf{Input}}\SetKwInOut{Output}{\textbf{Output}} % 替换关键词
    \Input{The discrete state space $\Gamma_{i,m}^S$ with $m\!=\!k,\dots,N_i$.}
    
    \Output{The optimal solution to the problem in (\ref{Eq.2}).}
    \BlankLine
   {Set $J_{i,N_i}^{*}(t_{i,N_i}^a)$ as in (\ref{Eq.14})\;}\vspace{3pt}
   
  \For{$m\!=\!N_i\!-\!1,\dots,k$}{\vspace{3pt} \For{$t_{i,m}^a\!\in\!\Gamma_{i,m}^S$}{\vspace{3pt} Compute $\Gamma_{i,m}^D(t_{i,m}^a)$ by (\ref{Eq.13})\;\vspace{3pt}
  \For{$t_{i,m}^w\!\in\!\Gamma_{i,m}^D(t_{i,m}^a)$}{\vspace{3pt} Compute $Q_{i,m}\big(t_{i,m}^a,t_{i,m}^w\big)$ by (\ref{Eq.12})\;}\vspace{3pt}
  Obtain $J_{i,m}^{*}\big(t_{i,m}^a\big)$ by (\ref{Eq.15})\;}}
  \vspace{3pt}
  
  \For{$m\!=\!k,\dots,N_i\!-\!1$}{Compute the optimal waiting time $t_{i,m}^{w*}=\arg\max\limits_{t_{i,m}^w\in\Gamma_{i,m}^D(t_{i,m}^a)}\!\!Q_{i,m}\big(t_{i,m}^a,t_{i,m}^w\big)$\;
  Compute the state 
  $t_{i,m+1}^a\!=\!f_{i,m}(t_{i,m}^a,t_{i,m}^{w*})$\;
  }\vspace{1pt}
  \textbf{return} the optimal waiting times $\{t_{i,k}^{w*},\dots,t_{i,N_i-1}^{w*}\}$.\vspace{1pt}
  \caption{DP-Based Optimal Solution}\label{Alg.2}
\end{algorithm}

\begin{figure*}[t]
     \centering
     \includegraphics[width=0.985\linewidth]{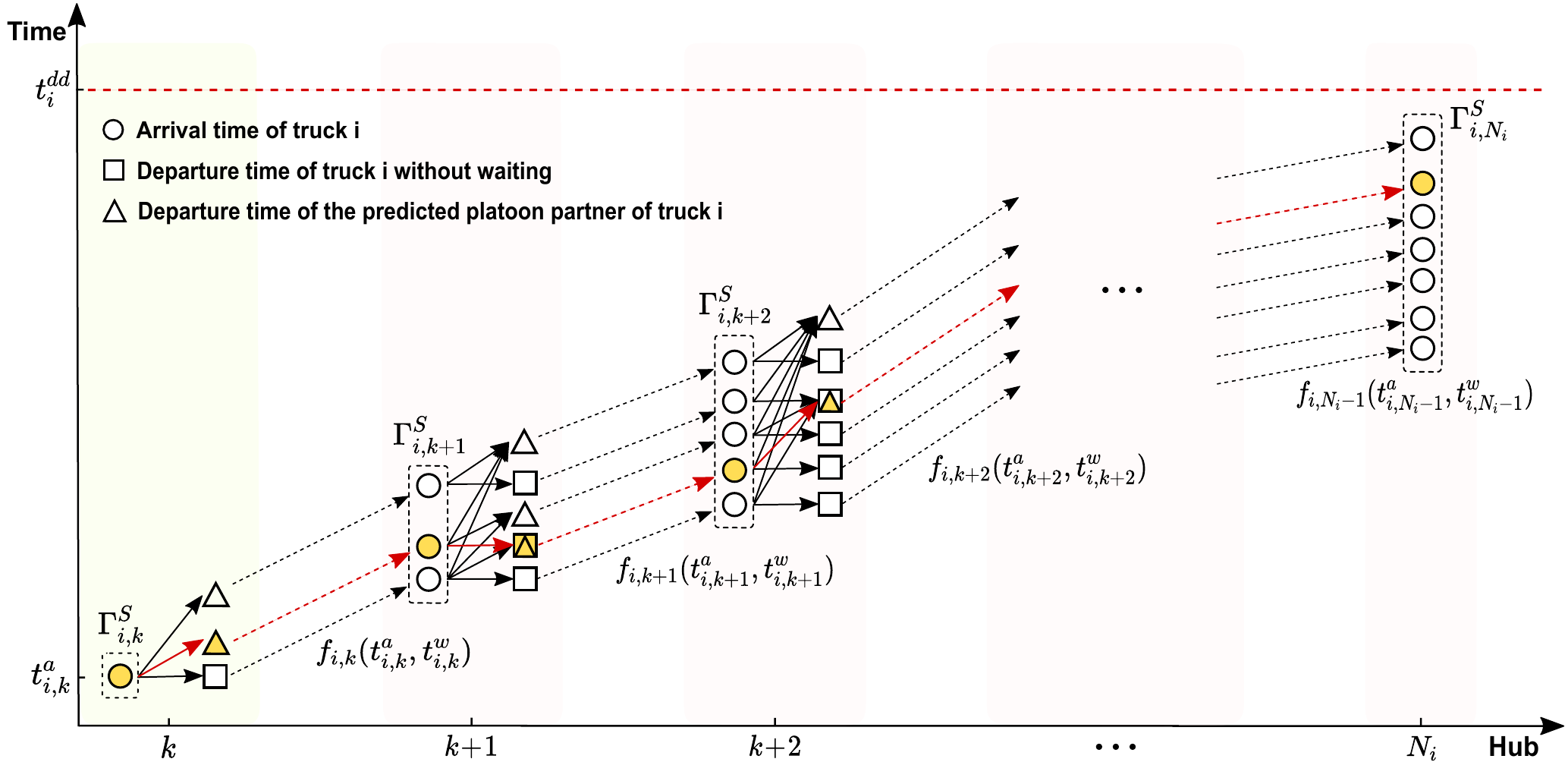}
     \vspace{-2pt}
      \caption{Illustration of the discrete decision and state spaces of truck $i$ that arrives at its $k$-th hub.}
      \label{Fig.3}
\end{figure*}

Based on the discrete decision and state spaces obtained above, the optimal solution to the problem (\ref{Eq.2}) is presented by the following Theorem~\ref{Theorem1}.  
\begin{theorem}\label{Theorem1}
The optimal solution to the problem in (\ref{Eq.2}) can be obtained by solving the following BOE. The optimal value function at the terminal stage is
\begin{align}
    J_{i,N_i}^{*}\big(t_{i,N_i}^a\big)=g_{i,N_i}(t_{i,N_i}^a), \quad ~\text{for all}~t_{i,N_i}^a\!\!\in\!\Gamma_{i,N_i}^S.\label{Eq.14}
\end{align}
The optimal value functions at stages $m\!=k,\dots,N_i\!-\!1$ are
\begin{align}
    &\!\!\!J_{i,m}^{*}\big(t_{i,m}^a\big)\nonumber\\
    &\!\!=\max\limits_{t_{i,m}^w\in\Gamma_{i,m}^D(t_{i,m}^a)}\!\!Q_{i,m}\big(t_{i,m}^a,t_{i,m}^w\big),\quad \text{for all $t_{i,m}^a\!\in\!\Gamma_{i,m}^S$},\label{Eq.15}
\end{align}
where $\Gamma_{i,m}^D(t_{i,m}^a)$ and $\Gamma_{i,m}^S$ are the discrete decision and state spaces computed by (\ref{Eq.13}) and Algorithm~\ref{Alg.1}, respectively.
\end{theorem}
\vspace{2pt}
\begin{proof} 
See the Appendix.
\end{proof}
\vspace{2pt}

The BOE in Theorem~\ref{Theorem1} can be solved by DP as in Algorithm~\ref{Alg.2}. The output of the algorithm is the optimal solution to the problem~(\ref{Eq.2}). We refer the readers to see, \emph{e.g.}, \cite{bellman1966dynamic,bertsekas2019reinforcement}, for detailed introductions on DP. 

In contrast to the conventional discretization methods using fixed time intervals to handle continuous state spaces, our method generates a discrete decision space $\Gamma_{i,m}^D(t_{i,m}^a)$ based on the discrete predicted departure times of other trucks. The constraint (\ref{Eq.3}) imposed by the delivery deadline and the conditions required to form platoons further contribute to the sparsity of the discretized decision space. As a result, the proposed approach is not subject to the curse of dimensionality associated with DP methods.  

\begin{remark}\label{Remark4}
Denote as $\tilde{n}\!=\!\max_{m\in\{1,\dots, N_i-1\}}|\Gamma_{i,m}^D|$ the maximum decision options of truck $i$ at each hub, where $\Gamma_{i,m}^D\!=\!\bigcup_{t_{i,m}^a\in\Gamma_{i,m}^S}\!\!\!\Gamma_{i,m}^D(t_{i,m}^a)$. The computational complexity of solving the problem (\ref{Eq.2}) by DP at the first hub of truck $i$ can be denoted as $O(\tilde{n}N_i)$, see, Example 1.3.1. in \cite{bertsekas2017dynamic}, where $\tilde{n}$ is no worse than $\max_{m\in\{1,\dots, N_i\}}\!\!\big|\Gamma_{i,m}^S\big|^2$.
\end{remark}

For a better understanding of the above results, we make use of Figure~\ref{Fig.3} to illustrate how one can generate the discrete decision space for truck $i$ that arrives at its $k$-th hub and obtain the DP graph associated with the problem (\ref{Eq.2}). By applying DP in such a DP graph, the optimal waiting time decisions of (\ref{Eq.2}) can be attained, as shown by the red arrows connecting the arrival and departure time nodes in orange in Figure~\ref{Fig.3}.

\begin{figure*}[!t]
     \centering
     \includegraphics[width=0.965\linewidth]{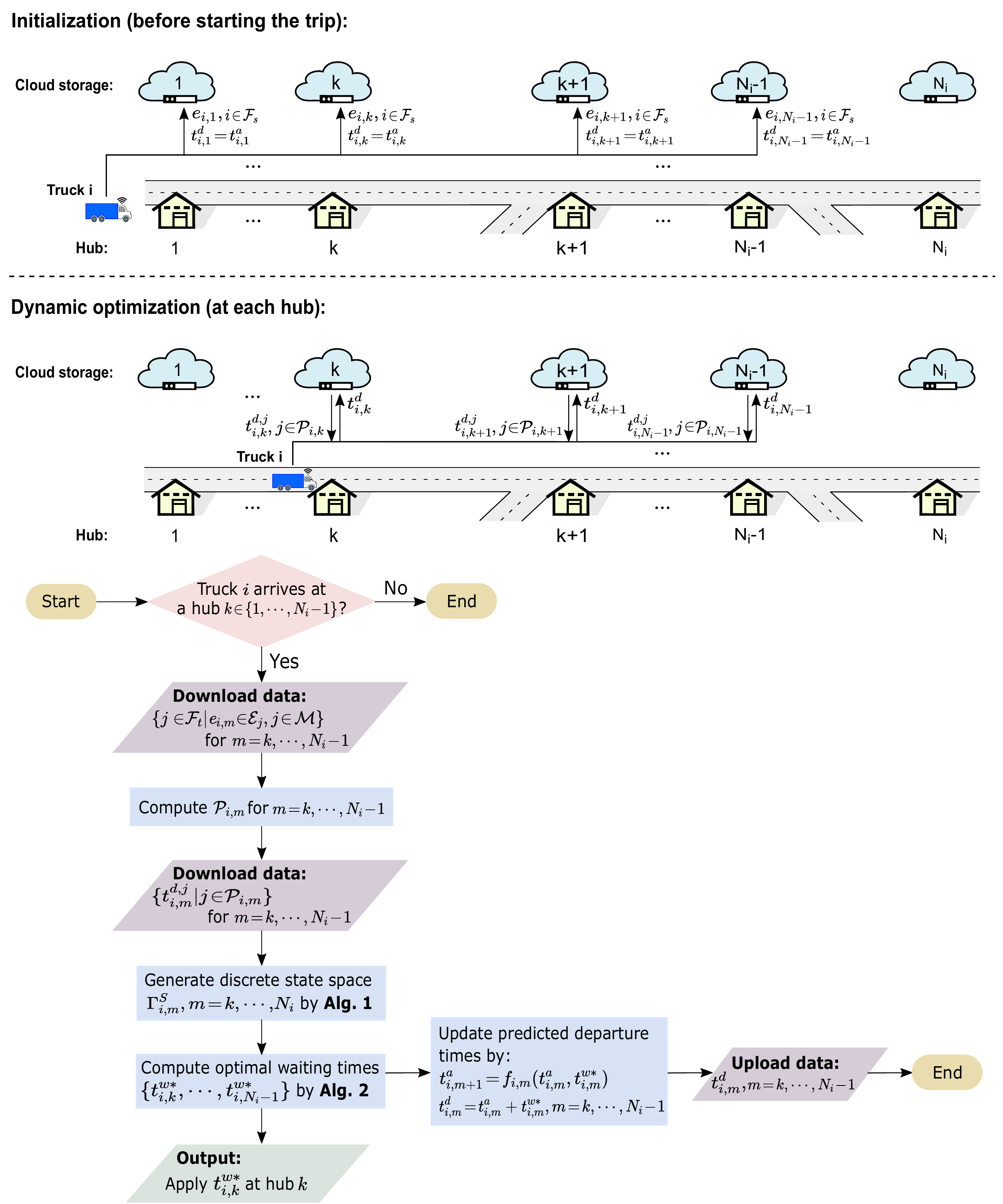}
      \vspace{-2pt}\caption{Illustration of the workflow applied to each truck in the platooning system.}
      \label{Fig.4}
\end{figure*}

The proposed platoon coordination method is applicable to large-scale transportation systems where hubs with cloud storage provide trucks with data storage and communication services while trucks conduct calculations independently for forming platoons. For any truck $i\!\in\!\mathcal{M}$, its workflow can be depicted in Figure~\ref{Fig.4} and described as follows:
\vspace{1pt}
\begin{itemize}
\item \textbf{Initialization:} Before starting the trip, truck $i$ uploads its route $e_{i,k}\!\in\!\mathcal{E}_i$, the fleet $i\!\in\!\mathcal{F}_s$ it belongs to, and its predicted departure times $t_{i,k}^d\!=\!t_{i,k}^a$ to the cloud storage for each hub $k$. At this stage, the predicted waiting time of truck $i$ at each hub is set to $0$. 
\vspace{2pt}

\item \textbf{Dynamic optimization:} When starting the trip, truck~$i$ is monitored at a regular time interval to check if it arrives at a hub $k\!\in\!\{1,\dots, N_i\!-\!1\}$. If arriving at a hub, truck $i$ downloads $\{j\!\in\!\mathcal{F}_t | \textbf{\textit{e}}_{i,m}\!\in\!\mathcal{E}_j, j\!\in\!\mathcal{M}\}$ from each of its hubs $m$ to compute its potential platoon partner-set $\mathcal{P}_{i,m}$ and downloads then the predicted departure times of other trucks $\{t_{i,m}^{d,j}|j\!\in\!\mathcal{P}_{i,m}\}$ with $m\!=\!k,\dots,N_i\!-\!1$. By applying Algorithms~\ref{Alg.1} and~\ref{Alg.2}, truck $i$ computes its optimal waiting times $\{t_{i,k}^{w*},\dots,t_{i,N_i-1}^{w*}\}$ and applies $t_{i,k}^{w*}$ at its hub $k$. Its predicted schedules at the following hubs are updated in accordance with $\{t_{i,k}^{w*},\dots,t_{i, N_i-1}^{w*}\}$.
\end{itemize}

\begin{table*}[t]
\caption{Fleet and Truck Assignment} % title name of the table
\vspace{-5pt}
\centering % centering table
\begin{tabular}{c|c|c|c|c|c|c|c}
  \hline\hline
  & & & & & &\\[-1.3ex]
  \raisebox{1.5ex}{\textbf{Fleet size}} & \raisebox{1.5ex}{\textbf{Nr. of trucks per fleet}} & \raisebox{1.5ex}{\textbf{Nr. of fleets}} &  \raisebox{1.5ex}{\textbf{Fleet index}} & \raisebox{1.5ex}{\textbf{Fleet percentage}} & \raisebox{1.5ex}{\textbf{Nr. of trucks}} & \raisebox{1.5ex}{\textbf{Truck index}} & \raisebox{1.5ex}{\textbf{Truck percentage}}\\
  [-1.1ex]
  \hline
  \multirow{3}*{Small fleet} & $1$ & $325$ & \multirow{3}*{$1$ -- $767$} & \multirow{3}*{$89.7\%$} & 325 &~ \multirow{3}*{$1-1971$} & \multirow{3}*{$39.4\%$}\\
   & 3 & 362 & ~ & ~ & 1086 & ~& ~\\
   & 7 & 80 & ~ & ~ & 560 &~ & ~\\
  \hline
  \multirow{3}*{Medium fleet} & 15 & 49 & \multirow{3}*{$768$ -- $851$}& \multirow{3}*{$9.8\%$} & 735 & \multirow{3}*{$1972$ -- $4216$}&\multirow{3}*{$44.9\%$}\\
  ~ & 34 & 27 & ~& ~ & 918 & ~&~\\
  ~ & 74 & 8 & ~& ~ & 592 & ~&~\\
  \hline
  \multirow{2}*{Large fleet} & 148 & 3 & \multirow{2}*{$852$ -- $855$} & \multirow{2}*{$0.5\%$} & 444 &\multirow{2}*{$4217$ -- $5000$} &\multirow{2}*{$15.7\%$}\\
  ~ & 340 & 1 & ~ & ~ & 340 &~ &~\\
  \hline
  \hline
\end{tabular}
\label{Table1}
\end{table*}

Note that our platoon coordination approach relies on two assumptions: (i) deterministic travel times in the dynamic model (\ref{Eq.1}), and (ii) free communication across different fleets to share routes and predicted schedules among trucks. As a result, the proposed method is limited to an ideal transportation network without travel time uncertainties and a reliable communication network without trust or privacy concerns between different fleets. Notwithstanding, travel time uncertainties can be incorporated with our method by allowing trucks to update their predicted schedules \textit{periodically} or in an event-triggered manner. In addition, the second assumption can be addressed by reaching data-sharing agreements among trucks or employing encrypted data transmission approaches. 

%==========Section V================
%------Simulation Study-------------
\section{Simulation Study}\label{Section V}
\begin{figure}[t]
     \centering
     \includegraphics[width=0.9\linewidth]{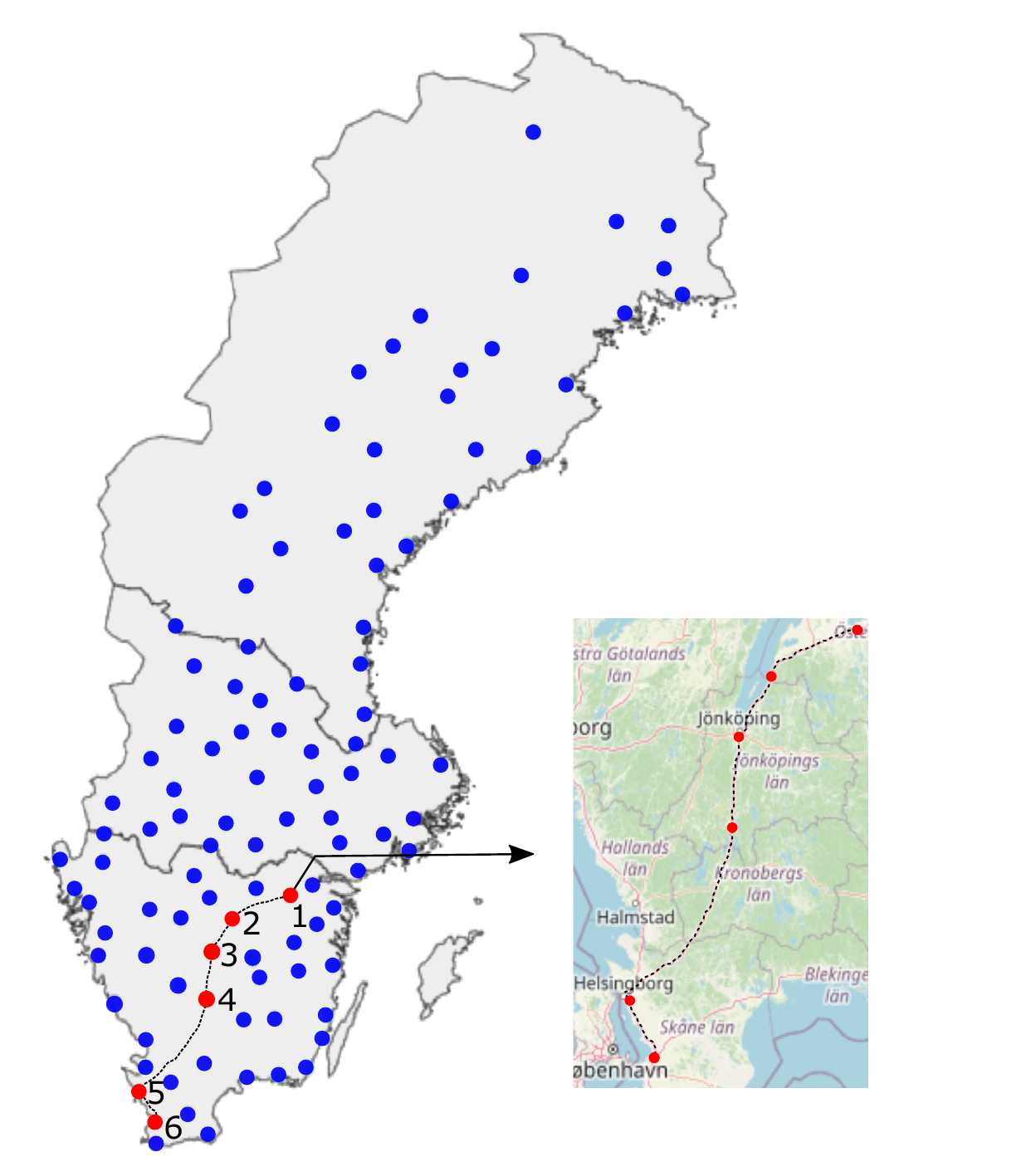}
     \vspace{-6pt}
      \caption{Swedish road network with $105$ hubs, where, as an illustrative example, hubs in the route of one truck are shown by the red nodes.}
      \label{Fig.5}
\end{figure}
In this section, we evaluate the improved platooning revenue from the developed multi-fleet platoon coordination approach in a large-scale transportation system. The simulation is conducted over the Swedish road network and based on the freight data in Sweden. In the following, the simulation setup and the simulation results will be introduced in detail.

%%%%%%%%%%%%%%%%%%%%%%%%%%%%%%%%%%%%%%%%%%
% Section V, Subsection A
\subsection{Simulation Setup}

% A-Sub-subsection 1)
\subsubsection{Road Network and Mission Generation} As shown in Figure~\ref{Fig.5}, we perform the simulation study in the Swedish road network with $105$ hubs, where each hub is a real road terminal in Sweden and represents the freight transport demands within one district. The coordinate of each hub is obtained from the SAMGODS model, which is the national model for freight transportation in Sweden and provides us with the truck flow between different districts based on the producer and consumer data. The mission of each truck is randomized such that its origin and destination belong to the set of hubs, and the truck flow from the SAMGODS model is used to get a realistic distribution of the transport missions. More specifically, the probability for two hubs $i$ and $j$ to be drawn as the origin and destination pair is computed by
\begin{align}
    P_{i,j}=\frac{F_{i,j}}{\sum_{ij}F_{i,j}},\nonumber
\end{align}
where $F_{i,j}$ is the truck flow from hub $i$ to hub $j$ in the SAMGODS model. The route between each pair of hubs is obtained from the open-source routing service \textit{OpenStreetMap}~\cite{OpenStreetMap}.

% A-Sub-subsection 2)
\begin{figure}[t]
     \centering
     \includegraphics[width=0.99\linewidth]{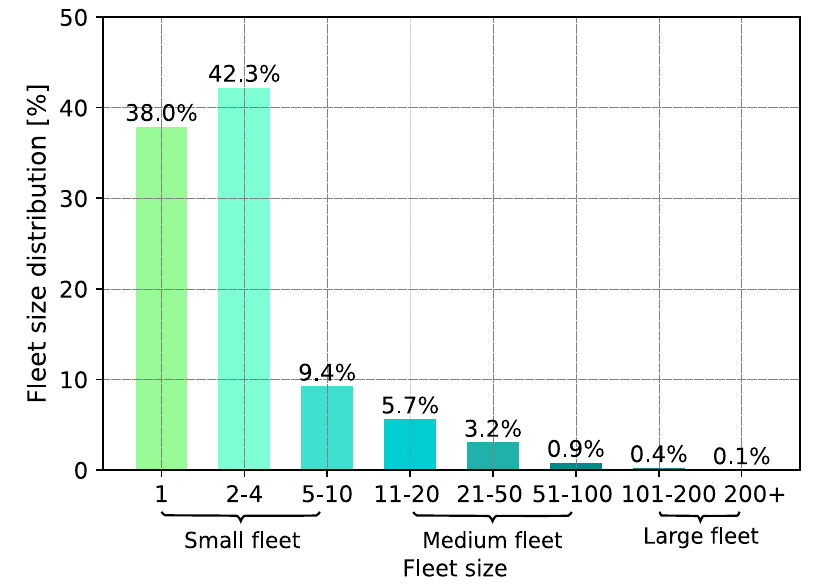}
     \vspace{-15pt}
      \caption{The fleet size distribution.}
      \label{Fig.6}
\end{figure}

\subsubsection{Fleet Distribution} We generate realistic fleet sizes from the data on the number of employees in transportation companies in Sweden \cite{TRAFA}. The fleet size distribution that we use in the simulation is given in Figure~\ref{Fig.6}, where the x-axis denotes the number of trucks in a fleet and the y-axis shows the fleet percentage. As is shown in the figure, $89.7\%$ of the fleets have no more than $10$ trucks in each fleet, and we refer to these fleets as small fleets. Fleets that include a number of trucks between $11$ and $100$, and over $100$ are referred to as medium and large fleets, respectively, which account for $9.8\%$ and $0.5\%$ in the total number of fleets.

According to the data reported in \cite{Lastbilstrafik}, approximately $5,000$ trucks start their transport trips each hour in Sweden. Following the fleet size distribution in Figure~\ref{Fig.6}, we can assign the $5,000$ trucks into $855$ fleets. More details about the assignment are provided in Table~\ref{Table1}.

% A-Sub-subsection 3)
\subsubsection{Parameter Settings}
We assume trucks start their trips at a random time between 08:00 to 09:00. The total travel time of a truck per day is less than $10$ hours, and the waiting budget (\emph{i.e.}, the maximum allowed waiting time at all hubs in the trip) of each truck is $10\%$ of its total travel time in the road network. In addition, we assume that trucks travel with a maximum and constant velocity of $80$ km per hour. The fuel consumption of each following truck in a platoon is assumed to be reduced by $10\%$, which leads to a monetary saving of $0.07$\texteuro ~per following truck per kilometer. Accordingly, the resulting platooning benefit $\xi$ is $5.6$\texteuro ~per following truck per hour. In line with the salary level of truck drivers in Sweden, the parameter $\epsilon_i$ representing the waiting loss of trucks is considered as $25$\texteuro ~per hour. 

%%%%%%%%%%%%%%%%%%%%%%%%%%%%%%%%%%%%%%
% Section V, Subsection B
\subsection{Solution Evaluation}
The simulation results and solution evaluation are introduced in this subsection. We conduct the simulation study over $5,000$ trucks belonging to $855$ different fleets and compare the platooning performance of three platoon coordination methods: the proposed predictive multi-fleet platooning method, the spontaneous multi-fleet platooning method, and the single-fleet platooning method. In spontaneous multi-fleet platooning, trucks communicate only with the trucks arriving at the same hub to optimize their schedules while having no access to the predicted schedules of other trucks at future hubs. The simulation results are evaluated from several perspectives, including the achieved reward, fuel savings, parameter sensitivity analysis, travel and waiting times, platooning rate and platoon formation rate, size of the formed platoons, and computation efficiency of the developed approach. For a better understanding of the dynamic optimization problem and the proposed solution scheme, the optimal scheduling solution of an illustrative example is first provided. 

\begin{table}[t]
\caption{The optimal schedule of one truck} % title name of the table
\vspace{-5pt}
\centering % centering table
\begin{tabular}{|c|c|c|c|c|} 
\hline
& & & &\\[-1.4ex]
  \raisebox{1.3ex}{\!\!\textbf{Hub}\!\!}& \raisebox{1.3ex}{\!\!\textbf{Arrival time}\!\!}&\raisebox{1.3ex}{\!\!\textbf{Departure time}\!\!} &\raisebox{1.3ex}{\!\!\textbf{Waiting time [s]}\!\!}&\raisebox{1.3ex}{\!\!\text{$|\mathcal{R}_{i,k}^{*}|$$\big/$$|\mathcal{P}_{i,k}|$\!\!}}
\\ [-0.5ex]
\hline % inserts single-line
& & & &\\[-1.1ex]
\raisebox{1.0ex}{\text{1}} & \raisebox{1.0ex}{\text{08:35:00}} & \raisebox{1.0ex}{\text{08:45:00}} & \raisebox{1.0ex}{\text{600}}& \raisebox{1.0ex}{\text{6}$\big/$\text{43}}\\[-0.4ex]
\hline 
& & & &\\[-1.1ex]
\raisebox{1.0ex}{\text{2}} & \raisebox{1.0ex}{\text{09:52:00}} & \raisebox{1.0ex}{\text{09:52:00}} & \raisebox{1.0ex}{\text{0}}& \raisebox{1.0ex}{\text{8}$\big/$\text{189}}
\\[-0.5ex]
\hline
& & & &\\[-1.1ex]
\raisebox{1.0ex}{\text{3}} & \raisebox{1.0ex}{\text{10:44:00}} & \raisebox{1.0ex}{\text{10:46:00}} & \raisebox{1.0ex}{\text{120}}& \raisebox{1.0ex}{\text{6}$\big/$\text{183}}
\\[-0.5ex]
\hline
& & & &\\[-1.1ex]
\raisebox{1.0ex}{\text{4}} & \raisebox{1.0ex}{\text{11:44:00}} & \raisebox{1.0ex}{\text{11:47:00}} & \raisebox{1.0ex}{\text{180}}& \raisebox{1.0ex}{\text{2}$\big/$\text{45}}
\\[-0.5ex]
\hline
& & & &\\[-1.1ex]
\raisebox{1.0ex}{\text{5}} & \raisebox{1.0ex}{\text{14:16:00}} & \raisebox{1.0ex}{\text{14:16:00}} & \raisebox{1.0ex}{\text{0}}& \raisebox{1.0ex}{\text{2}$\big/$\text{92}}
\\[-0.5ex]
\hline
& & & &\\[-0.9ex]
\raisebox{1.0ex}{\text{6}} & \raisebox{1.0ex}{\text{15:14:00}} & \raisebox{1.0ex}{/} & \raisebox{1.1ex}{/}& \raisebox{1.0ex}{/}\\[-0.1ex]
\hline
\end{tabular}
\label{Table2}
\end{table}

\begin{figure}[t]
     \centering
     \includegraphics[width=0.995\linewidth]{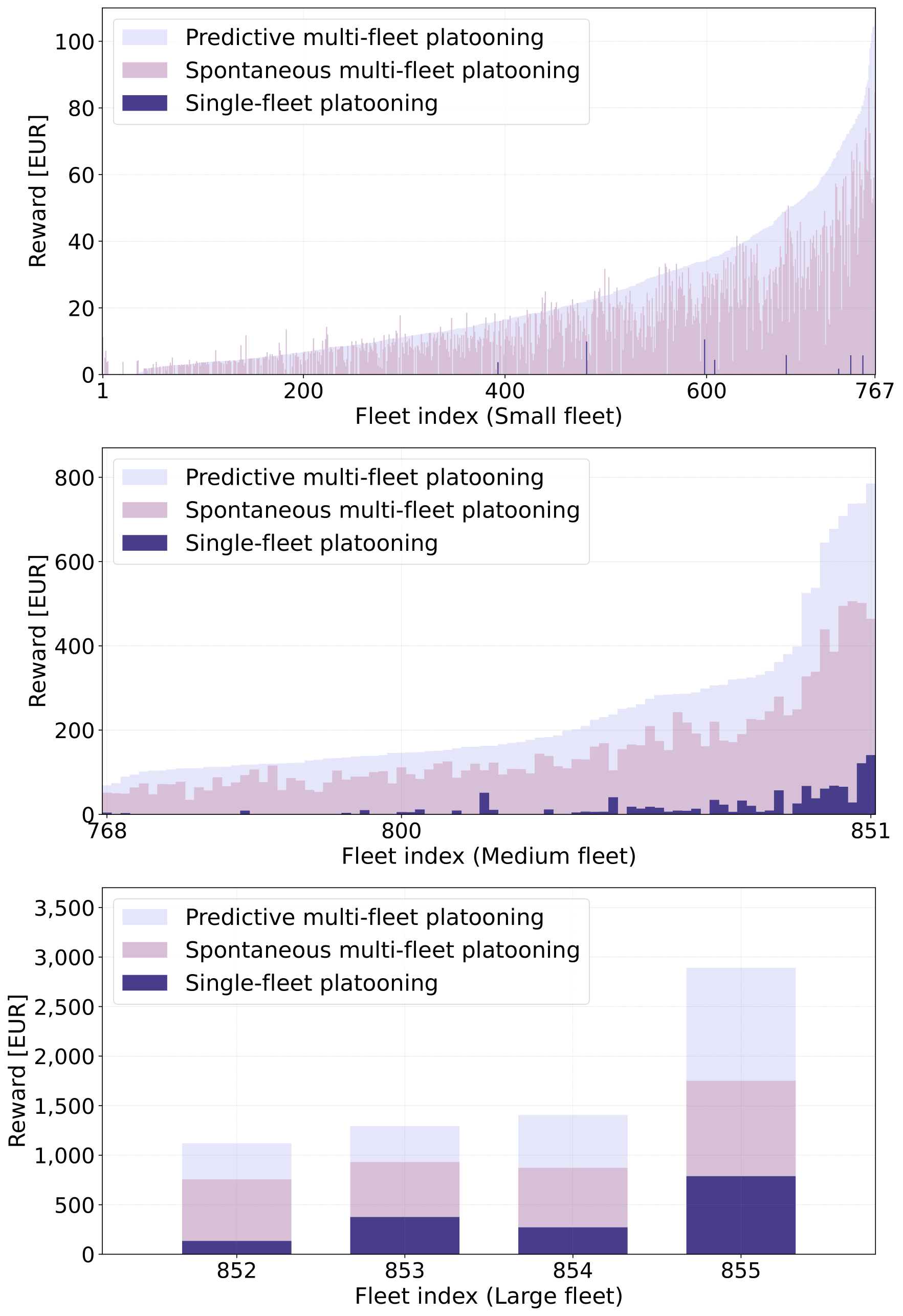}
     \vspace{-15pt}
      \caption{Reward of each small, medium, and large fleet.}
      \label{Fig.7}
\end{figure}
\begin{figure}[t]
     \centering
     \includegraphics[width=0.94\linewidth]{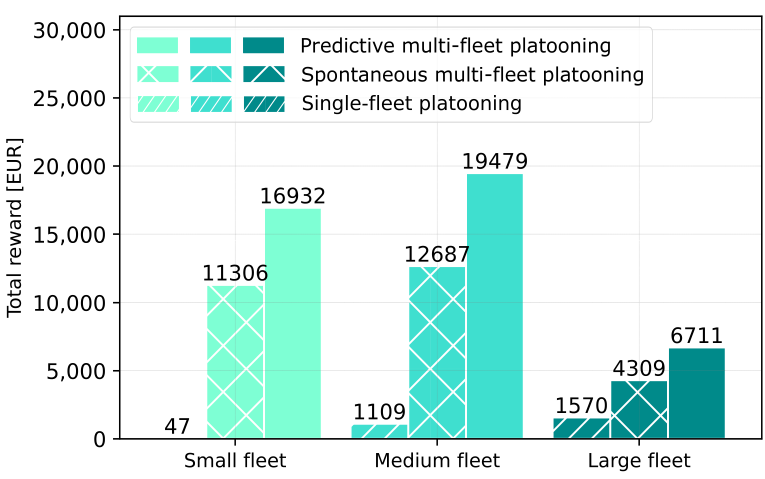}
     \vspace{-4pt}
      \caption{Total reward of small, medium, and large fleets.}
      \label{Fig.8}
\end{figure}

% B-Sub-subsection 1)
\subsubsection{Solution of One Truck} We provide in Table~\ref{Table2} the optimal schedule of one truck obtained by applying the predictive multi-fleet platoon coordination scheme. The transport route of the truck is given in Figure~\ref{Fig.5}. As shown in Table~\ref{Table2}, the truck starts its trip at $08\!:\!35\!:\!00$ and arrives at its destination hub at $15\!:\!14\!:\!00$. Its total travel time on roads is $384$ minutes, and the waiting budget at all hubs in its route is $38.4$ minutes. By the optimal schedule, $15$ minutes are spent at hubs for waiting and forming platoons. The last column of Table~\ref{Table2} gives the number of potential platoon partners of the truck, as well as that of its optimal platooning partners, at each hub.

% B-Sub-subsection 2)
\subsubsection{Reward}
Figure~\ref{Fig.7} shows the achieved reward (including the platooning reward and waiting loss) of each small, medium, and large fleet, compared among the three platooning schemes. In each sub-figure in Figure~\ref{Fig.7}, the fleet indices are sorted according to the fleet's reward achieved in the proposed predictive multi-fleet platooning. The simulation study shows that few small fleets form platoons and enjoy platooning benefits in single-fleet platooning, due to the small number of trucks included in each fleet. It also shows that the adoption of multi-fleet platoon coordination approaches leads to a substantial increase in fleet rewards. Additionally, it indicates that predictive multi-fleet platooning further improves the platooning profits based on the achievements of spontaneous multi-fleet platooning for each type of fleet.  
\begin{figure}[t]
     \centering
     \includegraphics[width=0.9\linewidth]{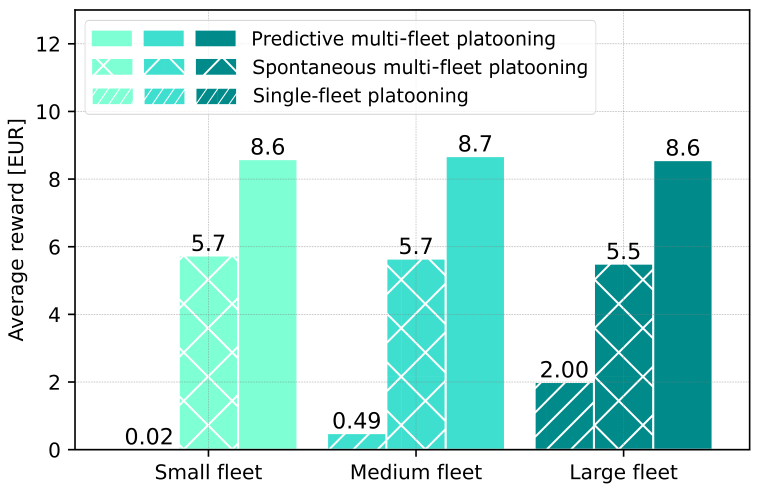}
      \caption{Average reward per truck in small, medium, and large fleets.}
      \label{Fig.9}
\end{figure}
\begin{figure}[t]
     \centering
     \includegraphics[width=0.9\linewidth]{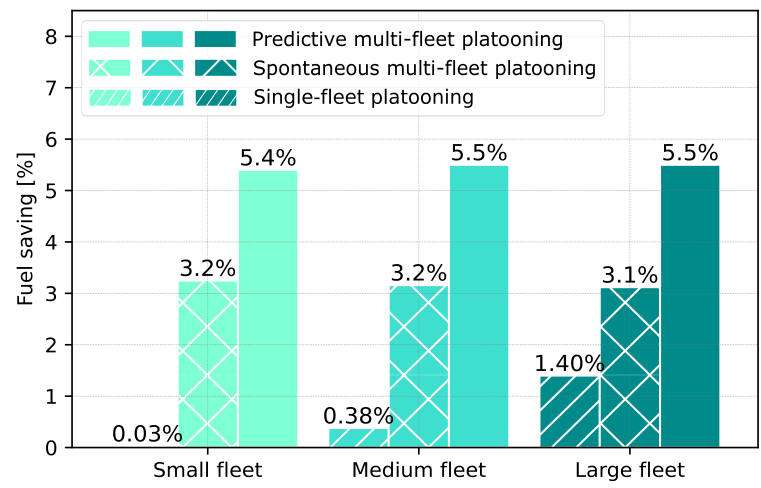}
      \caption{Fuel saving of small, medium, and large fleets.}
      \label{Fig.10}
\end{figure}
\begin{figure}[t]
     \centering
     \includegraphics[width=0.995\linewidth]{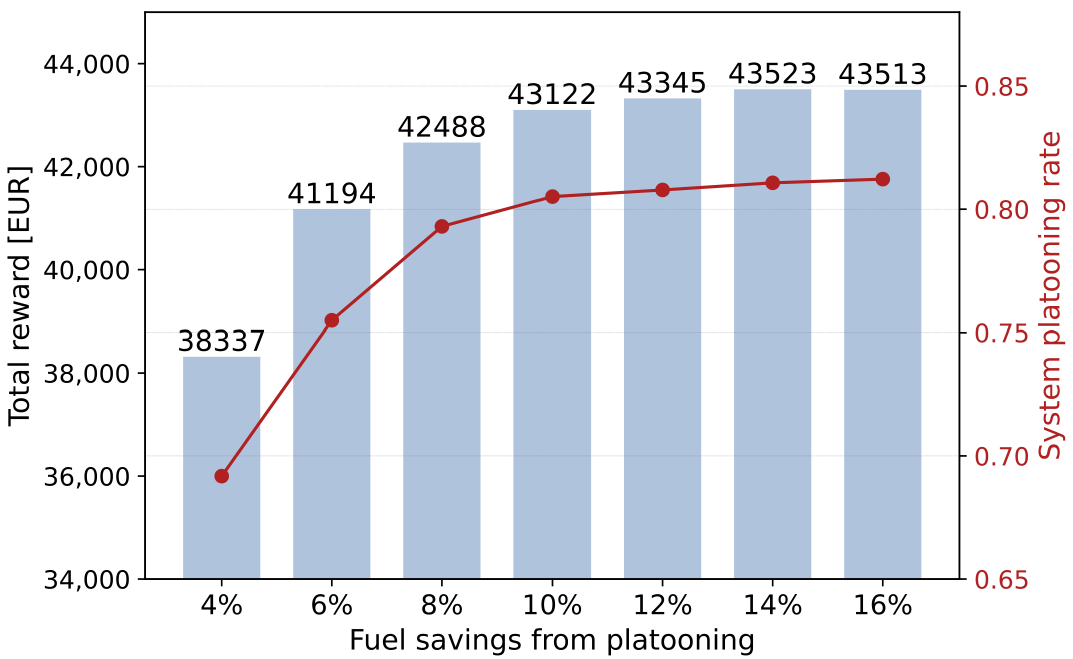}
     \vspace{-20pt}
      \caption{Total reward of all trucks in predictive multi-fleet platooning.}
      \label{Fig.11}
\end{figure}
\begin{figure}[t]
     \centering
     \includegraphics[width=0.905\linewidth]{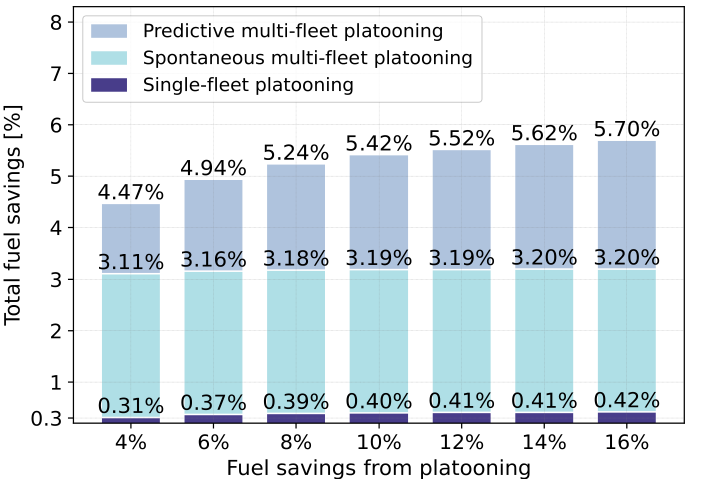}
     \vspace{-6.5pt}
      \caption{Total fuel savings of all trucks in three platooning methods.}
      \label{Fig.12}
\end{figure}

Figure~\ref{Fig.8} and Figure~\ref{Fig.9} show the total reward of trucks in small, medium, and large fleets, and the average reward per truck in each type of fleet, respectively. As we can see from Figure~\ref{Fig.8}, spontaneous multi-fleet platooning achieves higher monetary profits for small, medium, and large fleets compared to single-fleet platooning (approximately, $240$, $10$, and $1.7$ times higher, respectively). The predictive multi-fleet platooning further enhances these profits, with approximately $359$, $17$, and $3$ times higher profits for each fleet type, compared to single-fleet platooning. In total, the predictive multi-fleet platooning approach achieves around $15$ times higher monetary profit for all trucks in the system compared to single-fleet platooning, and results in about $0.5$ times higher monetary profit compared to spontaneous multi-fleet platooning. Although large fleets gain less increased reward compared to small fleets, they also have the incentive to cooperate with other fleets due to the substantial increase in their benefit. Furthermore, Figure~\ref{Fig.9} shows the average reward per truck in the three platoon coordination methods. The results indicate that the platooning profit gained by each truck in both the predictive and spontaneous multi-fleet platooning methods is independent of its fleet affiliation. 

% B-Sub-subsection 3)
\subsubsection{Fuel Saving}
\begin{figure*}[t]
\centering
\begin{minipage}{1\textwidth}
\centering
\subfigure[Travel time of individual trucks]
{\includegraphics[width=0.497\textwidth]{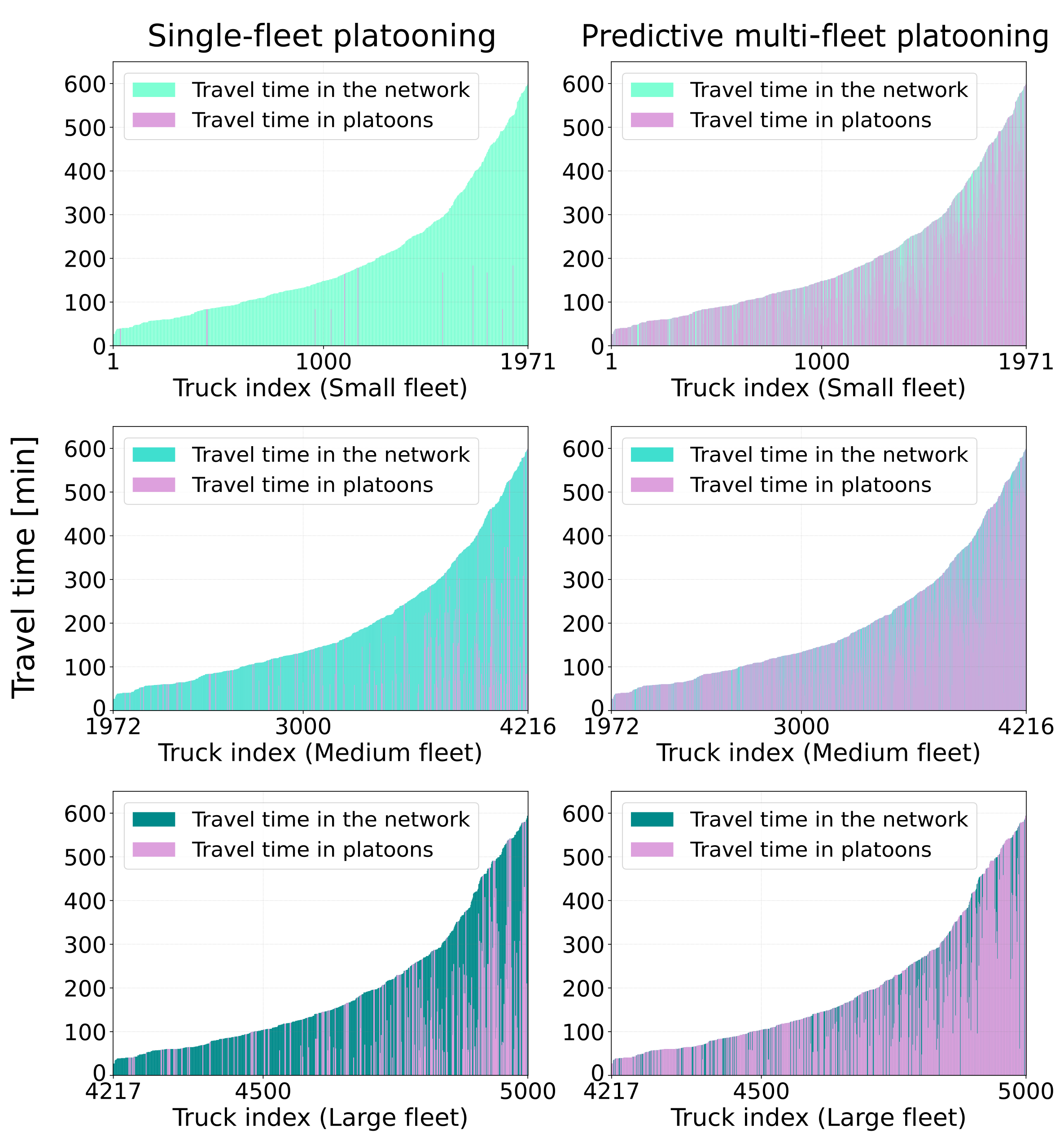}}
\subfigure[Waiting time of individual trucks]
{\includegraphics[width=0.497\textwidth]{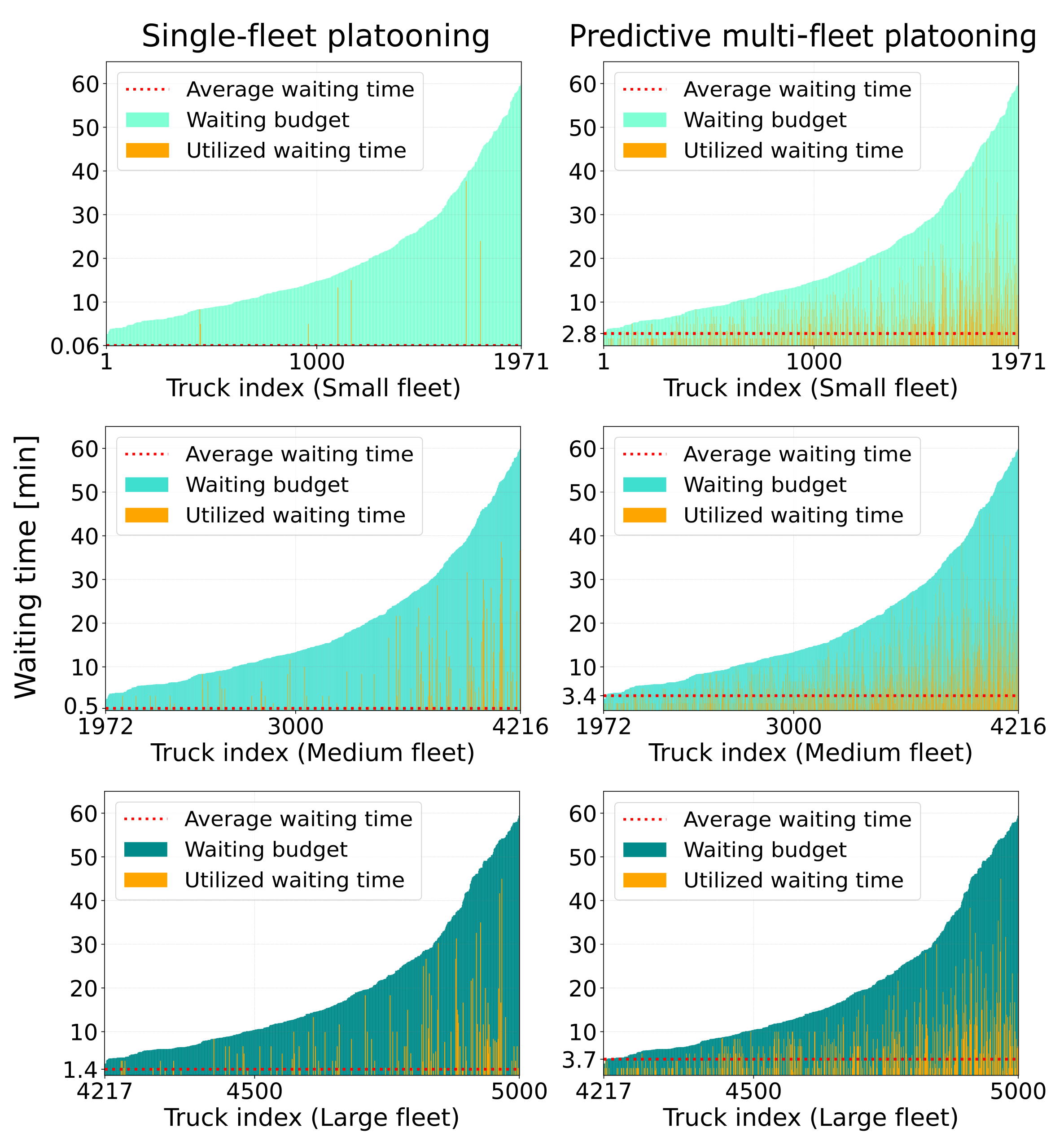}}
\end{minipage}
\vspace{-8pt}
\caption{(a) Trucks' travel times in platoons and in the road network, compared between single-fleet and predictive multi-fleet platooning. (b) The waiting budget and the utilized waiting time at all hubs in the route of individual trucks, compared between single-fleet and predictive multi-fleet platooning. The waiting budget of a truck equals $10\%$ of its total travel time in the road network.}
\label{Fig.13}
\end{figure*}
Figure~\ref{Fig.10} gives the total fuel savings of each type of fleet from platooning. The simulation results show that the proposed predictive multi-fleet platooning method achieves approximately $5.5\%$ fuel savings for small, medium, and large fleets, compared to $3.2\%$ in the spontaneous multi-fleet platooning scheme. In comparison with single-fleet platooning, where the fuel savings for small, medium, and large fleets are $0.03\%$, $0.38\%$, and $1.4\%$, respectively, the fuel economy is significantly improved by the predictive multi-fleet platooning. As a rough estimation, we utilize a linear model to convert fuel savings into reductions in CO$_2$ emissions, as adopted in~\cite{morrow2014assessment,xu2014energy}. Thereby, from a system perspective, the total fuel savings for all trucks achieved from predictive multi-fleet, spontaneous multi-fleet, and single-fleet platooning are $0.4\%$, $3.2\%$, and $5.5\%$, respectively. This corresponds to approximately $13$ times higher CO$_2$ emission reductions by employing the predictive multi-fleet platooning (\emph{i.e.}, $(5.5\!-\!0.4)/0.4\!=\!12.75$).

% B-Sub-subsection 4)
\subsubsection{Sensitivity Analysis}
Figure~\ref{Fig.11} shows how the fuel savings from truck platooning for the following trucks affect the trucks' total reward in the predictive multi-fleet platooning method. The y-axis shown on the right-hand side represents the platooning rate of the entire system, defined as the ratio of trucks' total travel time in platoons to that in the road network. The figure illustrates that as the percentage of fuel savings from platooning increases, the total reward tends to level off, indicating that the platooning rate of the system gradually approaches saturation status. Additionally, it shows that the total reward of the system drops sharply when the fuel-saving benefit of platooning becomes small. 

Figure~\ref{Fig.12} shows the total fuel savings achieved by all trucks in the platooning system applying the three platoon coordination methods. The results indicate that higher fuel savings from platooning lead to higher total fuel savings. Moreover, predictive multi-fleet platooning shows greater sensitivity to changes in platooning fuel savings compared to single-fleet and spontaneous multi-fleet platooning methods.

% B-Sub-subsection 5)
\subsubsection{Travel and Waiting Time}
The details of the travel and waiting times of individual trucks are provided in Figure~\ref{Fig.13}. Specifically, Figure~\ref{Fig.13}(a) gives the total travel time of each truck in the road network and in platoons from small, medium, and large fleets, compared between single-fleet and predictive multi-fleet platooning. In each sub-figure in Figure~\ref{Fig.13}(a), the x-axis represents trucks indices which are sorted according to trucks’ total travel times in the road network. By comparison, it can be seen that trucks' travel times in platoons increase significantly in the predictive multi-fleet platooning provided by our method for small, medium, and large fleets.  

Figure~\ref{Fig.13}(b) illustrates the waiting times of trucks at all hubs in their routes, where the waiting budget of each truck is assumed as $10\%$ of its total travel time in the road network. The truck index in the x-axis is consistent with the sorted truck indices in Figure~\ref{Fig.13}(a). As is shown, the average waiting time per truck in small, medium, and large fleets in the predictive multi-fleet platooning is $2.8$, $3.4$, and $3.7$ minutes, respectively. Compared to single-fleet platooning, there is an increase in the average waiting time for trucks in predictive multi-fleet platooning, while it leads to significant increases in trucks' travel times in platoons and the resulting platooning profits.

% B-Sub-subsection 6)
\subsubsection{Platooning Rate and Platoon Formation Rate} 
\begin{figure*}[t]
\centering
\begin{minipage}{1\textwidth}
\centering
\subfigure[Road transport flow]
{\includegraphics[width=0.329\textwidth]{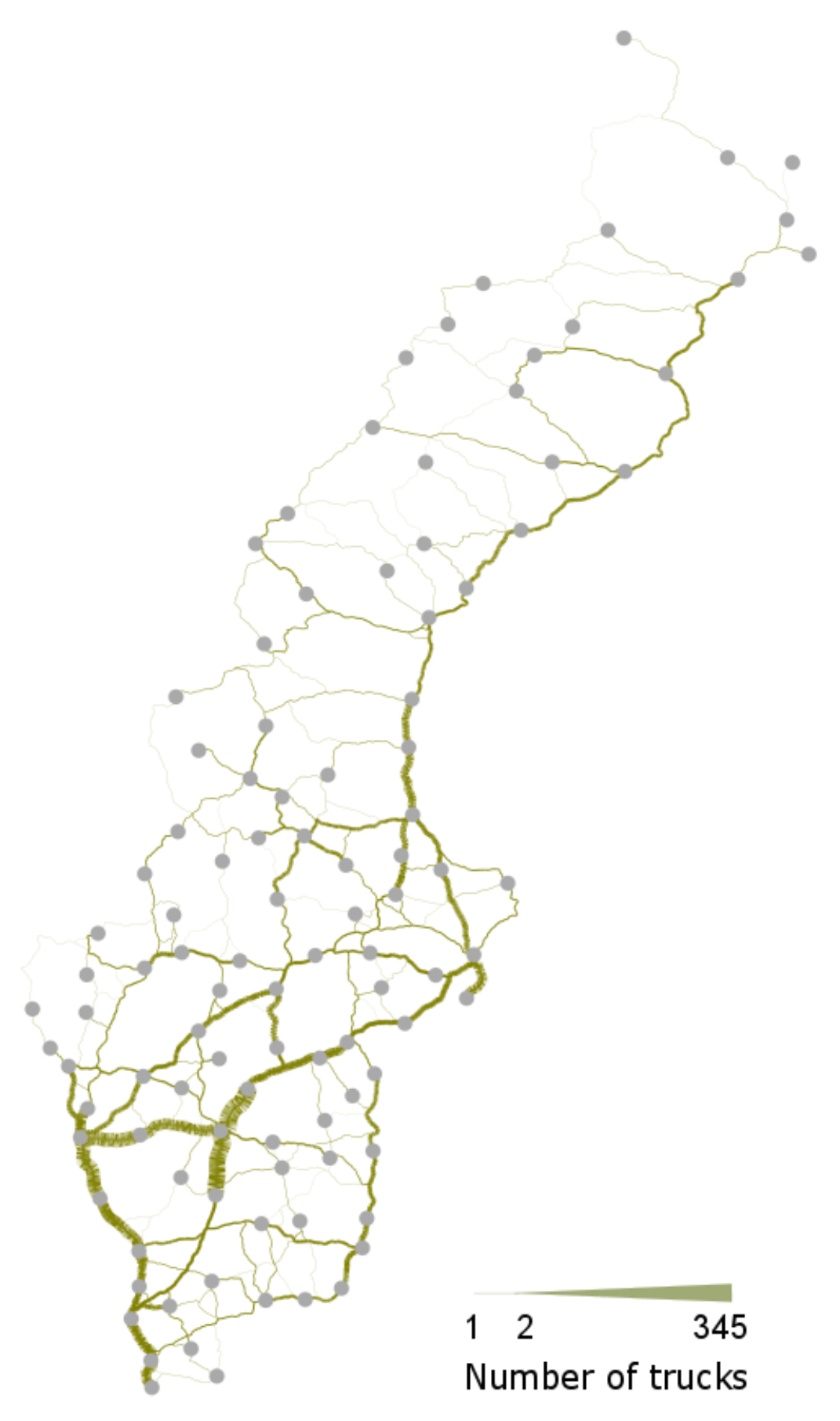}}
\subfigure[Platooning rate on roads]
{\includegraphics[width=0.329\textwidth]{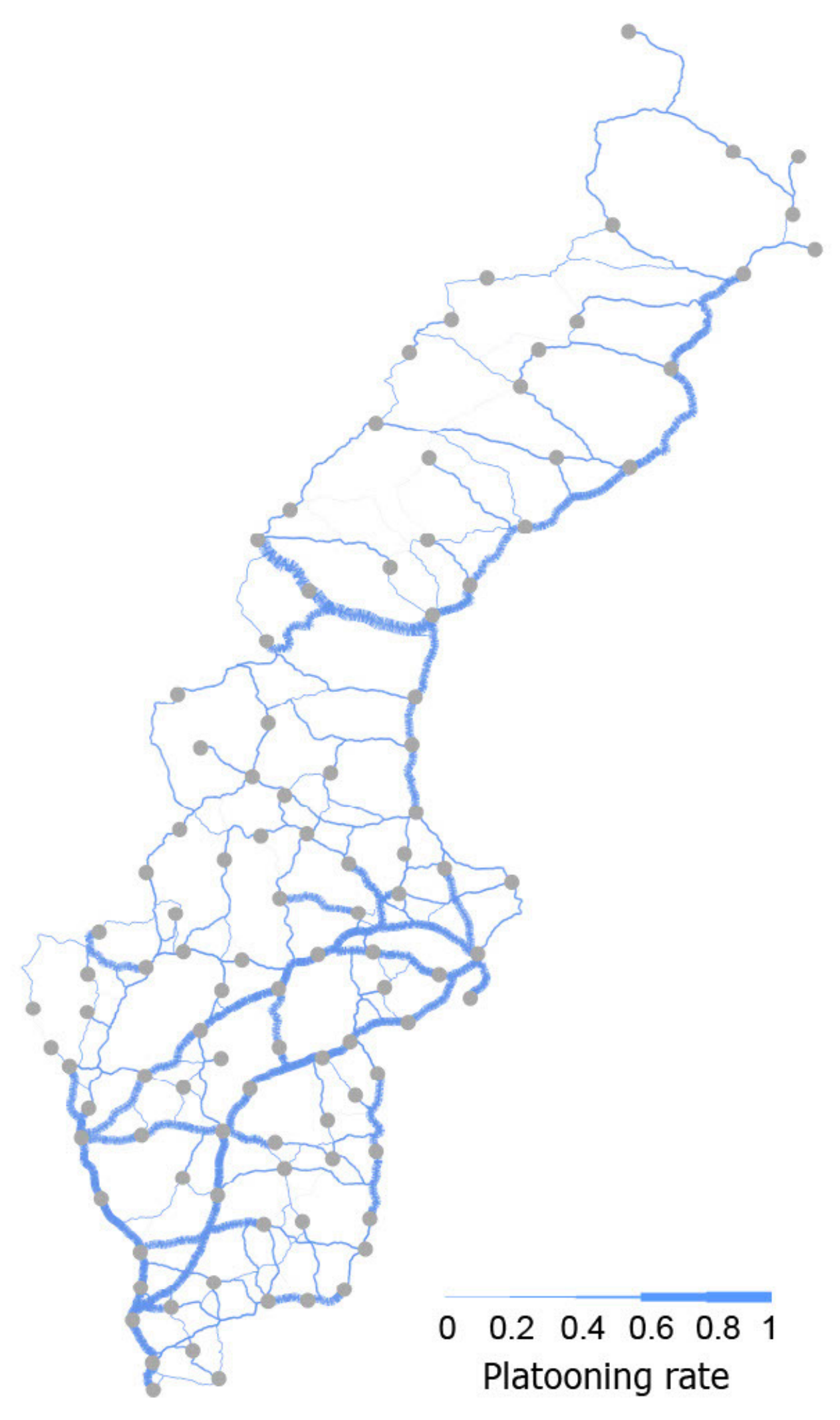}}
\subfigure[Platoon formation rate at hubs]
{\includegraphics[width=0.33\textwidth]{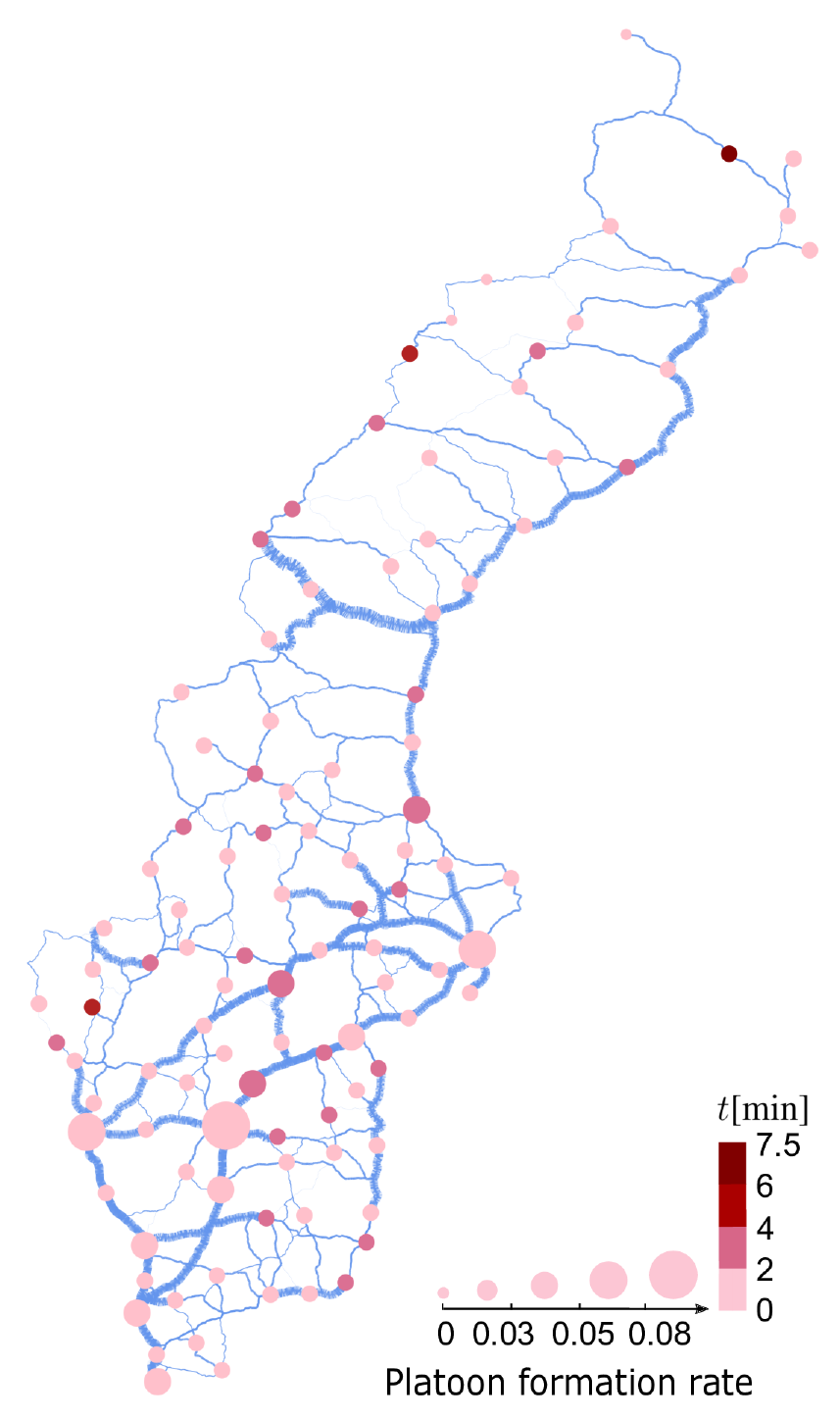}}
\end{minipage}
\caption{(a) The transport flow in the Swedish road network. (b) The platooning rate of each road segment in the predictive multi-fleet platooning method. (c) The platoon formation rate at hubs and trucks' average waiting time at hubs.}
\label{Fig.14}
\end{figure*}
We further evaluate trucks' platooning performance on roads and at hubs in the proposed predictive multi-fleet platoon coordination approach. As is shown in Figure~\ref{Fig.14}, the transport flow in the Swedish road network is given in sub-figure (a), which reflects the number of trucks traveling between different pairs of hubs per day in line with the SAMGODS data. Figure~\ref{Fig.14}(b) shows the platooning rate of all trucks in the system on each road segment in the road network, where the platooning rate on a road segment $\textbf{\textit{e}}\!\in\!\mathcal{E}$ is defined by
\begin{align}
    P_r(\textbf{\textit{e}})=\frac{\text{Nr. of following trucks on the road segment $\textbf{\textit{e}}$}}{\text{Nr. of trucks on the road segment $\textbf{\textit{e}}$}},\nonumber
\end{align}
where $P_r(\textbf{\textit{e}})$ takes values between $0$ and $1$. By comparing Figures~\ref{Fig.14}(a) and (b) one can see that trucks' platooning rate on a road segment is positively correlated with the transport flow on it. Moreover, Figure~\ref{Fig.14}(c) illustrates the platoon formation rate of each hub $k\!\in\!\mathcal{H}$ in the road network, which is defined as
\begin{align}
    P_{f}(k)\!=\!\frac{\text{Nr. of trucks finding new platoon partners at hub $k$}}{\text{Nr. of trucks in the road network}},\nonumber
\end{align}
where $P_f(k)$ measures each hub's contribution to the formation of platoons. For instance, a hub with a platoon formation rate of $0.04$ indicates that $4\%$ of trucks find new platoon partners at this hub. The higher the platoon formation rate of a hub, the more it contributes to forming platoons. We also show in Figure~\ref{Fig.14}(c) the trucks' average waiting time at each hub. The results show that hubs with a higher platoon formation rate have a relatively lower average waiting time. This is reasonable because hubs with high platoon formation rates generally also have a high throughput of trucks, as shown in Figure~\ref{Fig.14}(a), which leads to more platooning opportunities and a smaller time difference between truck arrivals. The evaluation results of hubs in Figure~\ref{Fig.14}(c) can be used to plan transport routes for trucks or to decide at which hubs to improve the capacity to facilitate more trucks in forming platoons. 
\begin{figure}[!t]
     \centering
     \includegraphics[width=0.85\linewidth]{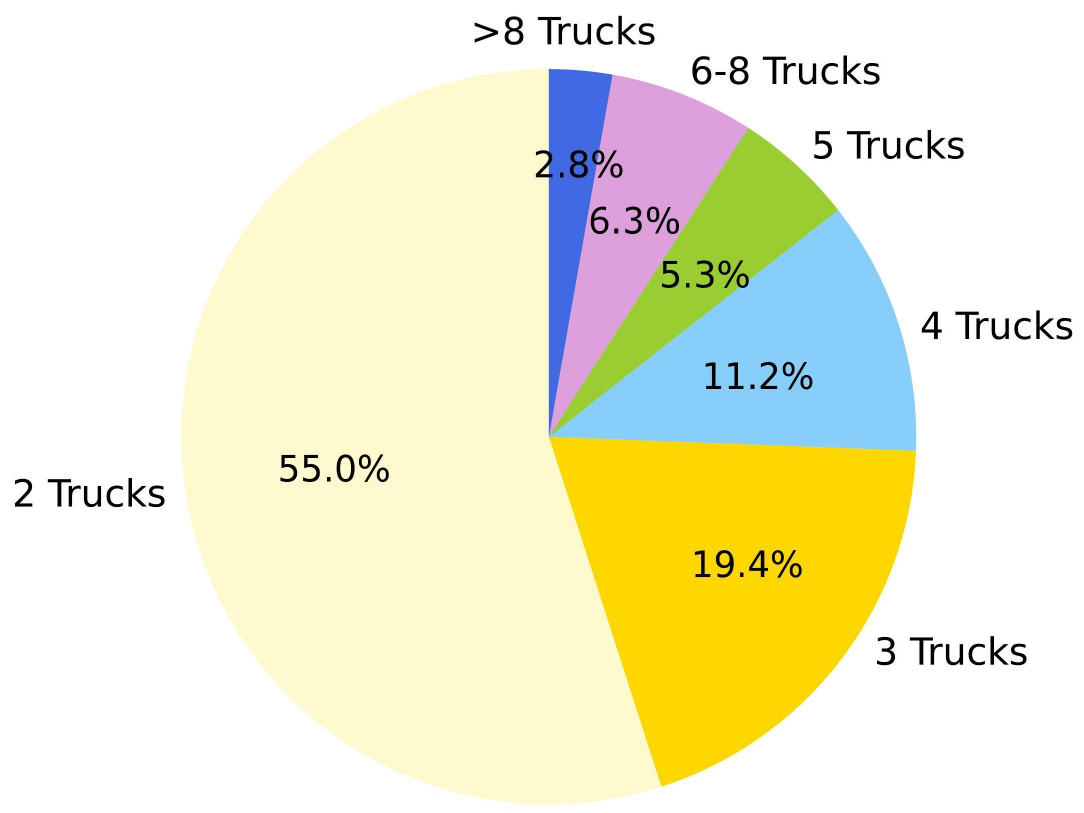}
     \vspace{-2pt}
      \caption{Size distribution of the formed platoons.}
      \label{Fig.15}
\end{figure}

% B-Sub-subsection 7)
\subsubsection{Platoon Size}
In our simulation study, $2453$ platoons are finally formed, and the size distribution of the formed platoons is shown in Figure~\ref{Fig.15}. As is shown, $55\%$ of the platoons consist of two trucks, $19.4\%$ and $11.2\%$ of the platoons are formed by $3$ and $4$ trucks, and around $97\%$ of the platoons have no more than $8$ trucks in the platoon. In practice, the platoon size might be limited due to safety concerns. Our method can be extended to handle platoon size constraints by limiting the size of the predicted platoon partner set or splitting the platoons exceeding the size limit.

% B-Sub-subsection 8)
\subsubsection{Computation Efficiency}
\begin{figure}[t]
     \centering
     \includegraphics[width=0.982\linewidth]{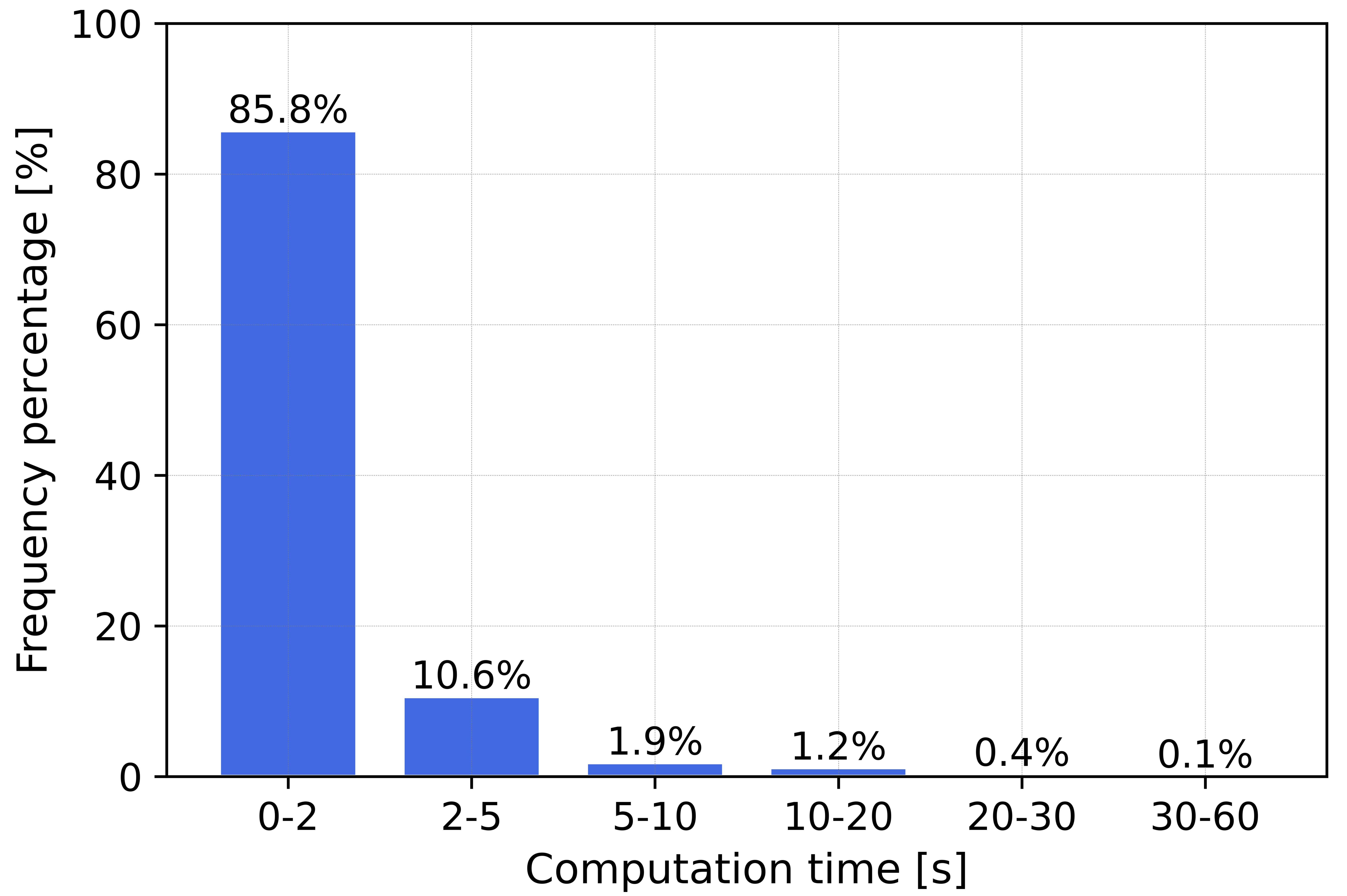}
     \vspace{-5pt}
      \caption{Computation time of trucks at each decision-making instance.}
      \label{Fig.16}
\end{figure}
\begin{table}[t]
\caption{Comparison of the computation efficiency} % title name of the table
\vspace{-5pt}
\centering % centering table
\begin{tabular}{|c|c|c|c|c|} 
\hline
& & & &\\[-1.4ex]
   \raisebox{1.3ex}{\!\!\textbf{Case}\!\!}&\raisebox{1.3ex}{\!\!\textbf{$\tilde{n}\!=\!\max_{m\in\{1,\dots,N_i\!-\!1\}}\!\!\big|\Gamma_{i,m}^D\big|$}\!\!}&\raisebox{1.3ex}{\textbf{$N_i$}} &\raisebox{1.3ex}{\!\textbf{Benchmark [s]}\!}&\raisebox{1.3ex}{\!\textbf{DP [s]}\!}
\\ [-0.5ex]
\hline % inserts single-line
& & & &\\[-1.0ex]
\raisebox{0.8ex}{\text{1}} & \raisebox{0.8ex}{\text{135}} & \raisebox{0.8ex}{\text{7}}& \raisebox{0.8ex} {\text{5.7}} &\raisebox{0.8ex}{\text{1.9}} \\[-0.5ex]
\hline 
& & & &\\[-1.0ex]
 \raisebox{0.8ex}{\text{2}} & \raisebox{0.8ex}{\text{283}} & \raisebox{0.8ex}{\text{6}}& \raisebox{0.8ex}{\text{45.9}} &\raisebox{0.8ex}{\text{4.8}}
\\[-0.5ex]
\hline
& & & &\\[-1.0ex]
 \raisebox{0.8ex}{\text{3}} & \raisebox{0.8ex}{\text{354}} & \raisebox{0.8ex}{\text{8}}& \raisebox{0.8ex}{\text{654.1}}&\raisebox{0.8ex}{\text{7.4}} 
\\[-0.5ex]
\hline
& & & &\\[-1.0ex]
\raisebox{0.8ex}{\text{4}} & \raisebox{0.8ex}{\text{387}} & \raisebox{0.8ex}{\text{7}}& \raisebox{0.8ex}{\text{1639.2}}&\raisebox{0.8ex}{\text{10.8}} 
\\[-0.5ex]
\hline
& & & &\\[-1.0ex]
 \raisebox{0.8ex}{\text{5}} & \raisebox{0.8ex}{\text{1319}} & \raisebox{0.8ex}{\text{7}}& \raisebox{0.8ex}{\text{7330.4}}&\raisebox{0.8ex}{\text{26.4}}
\\[-0.2ex]
\hline
\end{tabular}
\label{Table3}
\end{table}
Eventually, the computation efficiency of the proposed platoon coordination scheme is evaluated. We analyze the computation time that each truck takes to make its waiting time decision at every hub along its route and show the analysis results in Figure~\ref{Fig.16}. As we can see, over $98\%$ of the decision-making instances take less than $10$ seconds to compute the optimal waiting times, and more than $96\%$ of the decision-making instances take less than $5$ seconds. This demonstrates the high computation efficiency of the proposed method.

The computation efficiency of our solution scheme is further compared with a benchmark method, where all the possible combinations of the waiting time options at each hub are enumerated to obtain the optimal solution. We conduct simulation studies on $5$ trucks selected from the $5,000$ trucks, where each truck has a distinct route. For each truck (\textit{i.e.}, case), the computation time used to solve the predictive multi-fleet platoon coordination problem in DP and the benchmark solution method are given in Table~\ref{Table3}, where, as defined in Remark~\ref{Remark4}, $\tilde{n}$ denotes the maximum decision options of a truck at each hub. Based on our theoretical result, the optimal waiting time solution computed in the two methods is the same in each case, while the computation efficiency is significantly improved by DP. The above simulation studies demonstrate the high computation efficiency of the developed method, making our DP-based coordination approach suitable for real-time truck platooning in large transportation networks. 

%==========Section VI=================
\section{Conclusion}\label{Section VI}
This paper develops a DP-based platoon coordination method that schedules the waiting times of trucks in real-time and takes into account that trucks belong to different fleets interested in optimizing their own fleets' profits. A distributed dynamic optimization model is established to formulate the multi-fleet platoon coordination of trucks, where the reward function is modeled to capture the platooning reward and waiting loss of the fleet. Moreover, a DP-based optimal solution is presented for addressing the multi-fleet platoon coordination problem, where we show that the continuous decision space of the problem can be discretized without loss of optimality, which results in an efficient and real-time solution approach. Finally, the profit and efficiency of the platoon coordination method are demonstrated in a realistic simulation study performed over the Swedish road network, where we compare the case where trucks cooperate across fleets in forming platoons to the case where only single-fleet platoons are formed. 

The simulation study shows that multi-fleet platooning is essential to get significant economic and environmental benefits for the transportation system. Compared to single-fleet platooning, the developed multi-fleet platoon coordination approach achieves $359$, $17$, and $3$ times higher monetary profit for small, medium, and large fleets, respectively, and results in around $15$ times higher monetary profit for all trucks in the system. It also shows that compared to a system without any platooning, multi- and single-fleet platooning reduces the CO$_2$ emissions from $5,000$ trucks by $5.5\%$ and $0.4\%$, respectively. Moreover, the simulation study shows that over $98\%$ of the decision-making instances take a computation time of less than $10$ seconds, indicating that the proposed method is suitable for real-time platoon coordination in large transportation networks.

A limitation of the proposed method exists in the simplified truck dynamic model, where trucks' travel times are assumed to be deterministic, which in practice could be uncertain if considering traffic jams or various road conditions. In future work, we plan to extend the method in this paper to handle the travel time uncertainties by, for example, allowing trucks to update their predictions of arrival times at hubs. 

%==========Appendix=================
\appendix
\begin{prooflemma1}\label{proof_Lemma1}
Consider any waiting time $t_{i,m}^w\!\in\!{\Gamma_{i,m}(t_{i,m}^a)}\!\setminus\!{\Gamma_{i,m}^D(t_{i,m}^a)}$, meaning that $\Delta{f}\big(p_{i,m}^s,p_{i,m}^{-s}\big)\!=\!0$ and $t_{i,m}^w\!>\!0$. According to the stage reward function modeled in Eq.~(\ref{Eq.8}), we have that
\begin{align}
    \max\limits_{t_{i,m}^w\in{\Gamma_{i,m}(t_{i,m}^a)}\setminus{\Gamma_{i,m}^D(t_{i,m}^a)}}g_{i,m}(t_{i,m}^a,t_{i,m}^w)=0.\label{Eq.16}
\end{align}

Now consider another waiting time $t_{i,m}^{w'}\!\!=\!0\!\in\!\Gamma_{i,m}^D(t_{i,m}^a)$. By Eqs.~(\ref{Eq.7}) and (\ref{Eq.8}), the maximized stage reward function with respect to $t_{i,m}^a$ and $t_{i,m}^{w'}$ is denoted by
\begin{align}
    &\!\!\max\limits_{t_{i,m}^{w'}\in{\Gamma_{i,m}^D(t_{i,m}^a)}}g_{i,m}(t_{i,m}^a,t_{i,m}^{w'})\nonumber\\
    &\quad\quad\quad\quad\quad\quad~=\begin{cases}
        \!\Delta F_{i,m}^p(p_{i,m}^s,p_{i,m}^{-s})\!>\!0,& \text{if~$\exists\,t_{i,m}^{d,j}\!=\!t_{i,m}^a$}\\
        \!0, &\text{otherwise},
    \end{cases}\nonumber
\end{align}
where $t_{i,m}^{d,j}$ is the predicted departure time of other truck $j$ that belongs to the potential platoon partner set $\mathcal{P}_{i,m}$. Compared to the stage reward function in~(\ref{Eq.16}), it derives that
\begin{align}
    &\!\!\!\!\!\!\max\limits_{t_{i,m}^w\in{\Gamma_{i,m}(t_{i,m}^a)}\setminus{\Gamma_{i,m}^D(t_{i,m}^a)}}g_{i,m}(t_{i,m}^a,t_{i,m}^w)\nonumber\\
    &\quad \quad \quad \quad \quad \quad \quad \quad  \leq{\max\limits_{t_{i,m}^{w'}\in{\Gamma_{i,m}^D(t_{i,m}^a)}}g_{i,m}(t_{i,m}^a,t_{i,m}^{w'})}.\label{Eq.17}
\end{align}
Due to $t_{i,m}^{w'}\!<\!t_{i,m}^w$, and in line with the truck dynamics in (\ref{Eq.1}), the arrival time of truck $i$ at its next hub $(m\!+\!1)$ follows 
\begin{align}
    t_{i,m+1}^a>t_{i,m+1}^{a'},\label{Eq.18}
\end{align}
where $t_{i,m+1}^a\!=\!f_{i,m}(t_{i,m}^a,t_{i,m}^w)$ and $t_{i,m+1}^{a'}\!=\!f_{i,m}(t_{i,m}^a,t_{i,m}^{w'})$.
\vspace{-8pt}

Next, we will prove that $J_{i,m}^{*}\big(t_{i,m}^a\big)\!\leq\!{J_{i,m}^{*}\big(t_{i,m}^{a'}\big)}$ holds if there is $t_{i,m}^a\!>\!t_{i,m}^{a'}$, for $m\!=\!k,\dots,N_i$. We start with $m\!=\!N_i$.

\noindent (i) For $m\!=\!N_i$, by definition (\ref{Eq.10}), we have
\begin{align}
    J_{i,N_i}^{*}\big(t_{i,N_i}^a\big)\!=\!g_{i,N_i}(t_{i,N_i}^{a}),\quad \quad \text{for all}~t_{i,N_i}^a.\nonumber
\end{align}
If $t_{i,N_i}^a\!>\!t_{i,N_i}^{a'}$ holds, by Eq.~(\ref{Eq.9}), the following inequality holds
\begin{align}
    J_{i,N_i}^{*}\big(t_{i,N_i}^a\big)\!\leq\!J_{i,N_i}^{*}\big(t_{i,N_i}^{a'}\big).\label{Eq.19}
\end{align}

\noindent (ii) For $m\!=\!k,\dots,N_i\!-\!1$, the BOE in (\ref{Eq.11}) can be rewritten as
\begin{align}
    J_{i,m}^{*}\big(t_{i,m}^a\big)=\max\limits_{t_{i,m}^d\in\Gamma_{i,m}^d(t_{i,m}^a)}Q_{i,m}\big(t_{i,m}^a,(t_{i,m}^d\!\!-\!t_{i,m}^a)\big),\label{Eq.20}
\end{align}
where $t_{i,m}^d\!=t_{i,m}^a\!+t_{i,m}^w$ represents the departure time of truck $i$ at its $m$-th hub. The BOE in Eq.~(\ref{Eq.20}) shows that optimizing truck $i$'s waiting time $t_{i,m}^w$ at its $m$-th hub is equivalent to optimizing its departure time $t_{i,m}^d$ at the same hub. By (\ref{Eq.3}), $t_{i,m}^d$ is constrained by the set
\begin{align}
    \!\!\!\Gamma_{i,m}^d(t_{i,m}^a)\!=\!\Bigg\{t_{i,m}^d\,\bigg|\,t_{i,m}^a\!\leq\!{t_{i,m}^d}\!\leq\!{t_{i}^{dd}-\!\sum_{m=k}^{N_i-1}\tau(\textbf{\textit{e}}_{i,m})}\Bigg\}.\label{Eq.21}
\end{align}
So given two arrival times $t_{i,m}^a$ and $t_{i,m}^{a'}$, if $t_{i,m}^a\!>\!t_{i,m}^{a'}$ holds, by (\ref{Eq.21}), we have that
\begin{align}
    \Gamma_{i,m}^d(t_{i,m}^a)\!\subset\!{\Gamma_{i,m}^d(t_{i,m}^{a'})},\nonumber
\end{align}
which means that the later arrival time will cause a shrank decision space to the BOE in (\ref{Eq.20}), and it then will not result in a larger optimal reward function, \emph{i.e.}, 
\begin{align}
    J_{i,m}^{*}\big(t_{i,m}^a\big)\!\leq\!{J_{i,m}^{*}\big(t_{i,m}^{a'}\big)}.\label{Eq.22}
\end{align}

Given the above, if (\ref{Eq.18}) holds, there is
\begin{align}
    J_{i,m+1}^{*}\big(t_{i,m+1}^a\big)\!\leq\!{J_{i,m+1}^{*}\big(t_{i,m+1}^{a'}\big)}.\label{Eq.23}
\end{align}
Thus, by Eqs.~(\ref{Eq.11}), (\ref{Eq.12}), (\ref{Eq.17}) and (\ref{Eq.23}), it derives that 
\begin{align}
    &\!\max\limits_{t_{i,m}^w\in{\Gamma_{i,m}(t_{i,m}^a)}\setminus{\Gamma_{i,m}^D(t_{i,m}^a)}}\!\Big[g_{i,m}(t_{i,m}^a,t_{i,m}^w)\!+\!J_{i,m+1}^{*}\big(t_{i,m+1}^a\big)\Big]\nonumber\\
    &\quad \quad \quad~  \leq{\!\max\limits_{t_{i,m}^{w'}\in{\Gamma_{i,m}^D(t_{i,m}^a)}}\!\Big[g_{i,m}(t_{i,m}^a,t_{i,m}^{w'})\!+\!J_{i,m+1}^{*}\big(t_{i,m+1}^{a'}\big)\Big]}.\nonumber
\end{align}
Consequently, it proves that
\begin{align}
    J_{i,m}^{*}\big(t_{i,m}^a\big)&=\!\max\limits_{t_{i,m}^w\in{\Gamma_{i,m}(t_{i,m}^a)}}Q_{i,m}\big(t_{i,m}^a,t_{i,m}^w\big)\nonumber\\
    &={\!\max\limits_{t_{i,m}^w\in{\Gamma_{i,m}^D(t_{i,m}^a)}}Q_{i,m}\big(t_{i,m}^a,t_{i,m}^w\big)}.\nonumber
\end{align}
\end{prooflemma1}

\begin{proof}
The proof of Theorem~\ref{Theorem1} follows directly the proof of Lemma~\ref{Lemma1}, where we prove that the continuous decision space $\Gamma_{i,m}(t_{i,m}^a)$ of the BOE in (\ref{Eq.11}) can be discretized as $\Gamma_{i,m}^D(t_{i,m}^a)$, without loss of optimality. The discrete decision space $\Gamma_{i,m}^D(t_{i,m}^a)$ and the state transition function $f_{i,m}(t_{i,m}^a,t_{i,m}^w)$ lead to a discrete state space $\Gamma_{i,m}^S$. Thereby, it proves that the BOE in (\ref{Eq.11}) can be solved by DP using the discrete decision and state spaces generated by Lemma~\ref{Lemma1} and Algorithm~\ref{Alg.1}.
\end{proof}
\vspace{2pt}

\section*{Acknowledgment}
The authors would like to thank Albin Engholm for providing the simulation data from the SAMGODS model. The author Ting Bai would also like to thank the support of the Outstanding Ph.D. Graduate Development Scholarship from Shanghai Jiao Tong University, Shanghai, China.

\bibliographystyle{IEEEtran}
\bibliography{Ref}

\begin{IEEEbiography}[{\includegraphics[width=1in,height=1.25in,clip,keepaspectratio]{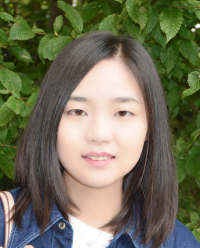}}]{\textcolor{black}{Ting Bai}}\textcolor{black}{received the B.Sc. degree in Automation from Northwestern Polytechnical University, Xi'an, China, in 2013, and the Ph.D. degree in Electrical Engineering from Shanghai Jiao Tong University, Shanghai, China, in 2019. Since 2020, she has been with the Division of Decision and Control Systems, Department of Electrical Engineering and Computer Science, KTH Royal Institute of Technology, Stockholm, Sweden, where she is a postdoctoral researcher. Her research interests include distributed model predictive control, dynamic programming methods, and their applications to transportation systems.} 
\end{IEEEbiography}
\vspace{-120pt}

\begin{IEEEbiography}[{\includegraphics[width=1in,height=1.25in,clip,keepaspectratio]{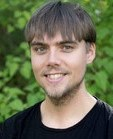}}]{Alexander Johansson} received a B.Sc. degree in vehicle engineering in 2015 and an M.Sc. degree in applied mathematics in 2017 from KTH Royal Institute of Technology, Stockholm, Sweden. In 2022, he received a Ph.D. from the Division and Control Systems, Department of Intelligent Systems, School of Electrical Engineering, KTH Royal Institute of Technology, Stockholm, Sweden. His research interests are optimization, game theory, and control for platoon coordination and swarm technology. 
\end{IEEEbiography}
\vspace{-120pt}

\begin{IEEEbiography}[{\includegraphics[width=1in,height=1.25in,clip,keepaspectratio]{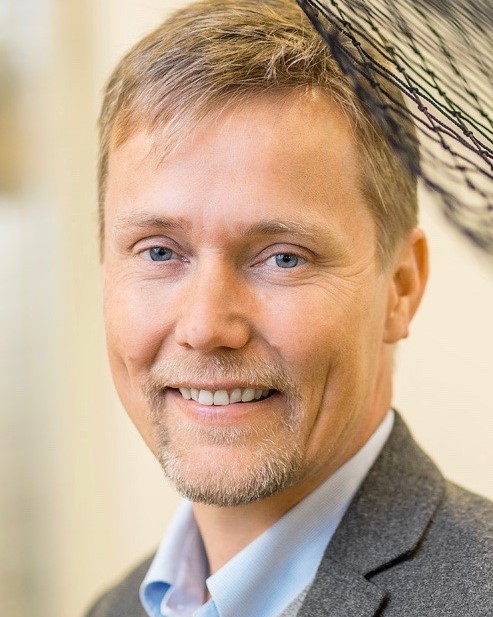}}]{Karl Henrik Johansson} is a Professor with the School of Electrical Engineering and Computer Science at KTH Royal Institute of Technology in Sweden and Director of Digital Futures. He received M.Sc. and Ph.D. degrees from Lund University. He has held visiting positions at UC Berkeley, Caltech, NTU, HKUST Institute of Advanced Studies, and NTNU. His research interests are in networked control systems and cyber-physical systems with applications in transportation, energy, and automation networks. He is a member of the Swedish Research Council's Scientific Council for Natural Sciences and Engineering Sciences. He has served on the IEEE Control Systems Society Board of Governors, the IFAC Executive Board, and is currently President of the European Control Association. He has received several best paper awards and other distinctions from IEEE, IFAC, and ACM. He has been awarded Distinguished Professor with the Swedish Research Council and Wallenberg Scholar with the Knut and Alice Wallenberg Foundation. He has received the Future Research Leader Award from the Swedish Foundation for Strategic Research and the triennial Young Author Prize from IFAC. He is Fellow of the IEEE and the Royal Swedish Academy of Engineering Sciences, and he is IEEE Control Systems Society Distinguished Lecturer.
\end{IEEEbiography}
\vspace{-120pt}

\begin{IEEEbiography}[{\includegraphics[width=1in,height=1.25in,clip,keepaspectratio]{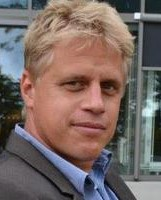}}]{Jonas M{\aa}rtensson} received the M.Sc. degree in vehicle engineering and the Ph.D. degree in automatic control from KTH Royal Institute of Technology, Stockholm, Sweden, in 2002 and 2007, respectively. In 2016, he was appointed as a Docent. He is currently a Professor with the Division of Decision and Control Systems, KTH Royal Institute of Technology. He is also engaged as the Director of the Integrated Transport Research Laboratory and the Thematic Leader for the area transport in the information age with the KTH Transport Platform. His research interests are cooperative and autonomous transport systems, in particular related to heavy-duty vehicle platooning. He is involved in several collaboration projects with Scania CV AB, S{\"o}dert{\"a}lje, Sweden, dealing with collaborative adaptive cruise control, look-ahead platooning, route optimization and coordination for platooning, path planning and predictive control of autonomous heavy vehicles, and related topics.
\end{IEEEbiography}
\end{document}